%% file: main.tex
\documentclass[journal]{IEEEtran}
\usepackage{microtype}

\usepackage{cite}

\ifCLASSINFOpdf
\else
\fi
\usepackage{amsmath}
\usepackage{amsfonts}
\usepackage{bbm}
\usepackage{amsthm}
\usepackage{mathtools}

\usepackage{xcolor}
\usepackage{mymacros}
\usepackage{tikz}
\usetikzlibrary{arrows,positioning,calc,external,backgrounds}
\usepackage{pgfplots}
\usepackage{amssymb}
\pgfplotsset{compat=newest}

\begin{document}

%
\title{Orthogonal Time Frequency Space Modulation: \newline A Discrete Zak Transform Approach}
%
%

\author{Franz~Lampel,~\IEEEmembership{Student Member,~IEEE,}
        Alex~Alvarado,~\IEEEmembership{Senior Member,~IEEE,}
        and~Frans~M.J.~Willems,~\IEEEmembership{Fellow,~IEEE
        }
\thanks{The authors are with the Information and Communication  Theory Lab,  Signal  Processing  Systems Group,   Department   of   Electrical   Engineering,   TU/e,   5600   Eindhoven, The Netherlands   (e-mails: \{f.lampel, a.alvarado, f.m.j.willems\}@tue.nl).}
\thanks{This research is supported by the Dutch Technology Foundation TTW, which is part of the Netherlands Organisation for Scientific Research (NWO), and which is partly funded by the Ministry of Economic Affairs under the project
Integrated Cooperative Automated Vehicles (i-CAVE).}
}

%
%

\markboth{Preprint, \today}{}%
%



\maketitle

\begin{abstract}
In orthogonal time frequency space (OTFS) modulation, information-carrying symbols reside in the delay-Doppler (DD) domain. By operating in the DD domain, an appealing property for communication arises: time-frequency (TF) dispersive channels encountered in high mobility environments become time-invariant. The time-invariance of the channel in the DD domain enables efficient equalizers for time-frequency dispersive channels. In this paper, we propose an OTFS system based on the discrete Zak transform. The presented formulation not only allows an efficient implementation of OTFS but also simplifies the derivation and analysis of the input-output relation of TF dispersive channel in the DD domain. 
\end{abstract}

\begin{IEEEkeywords}
Discrete Zak transform, discrete Fourier transform, doubly dispersive channel, mmWave, OFDM, OTFS, time-frequency dispersive channel, Zak transform.
\end{IEEEkeywords}

%
\IEEEpeerreviewmaketitle

\section{Introduction}
\input{Introduction}
%
%
%
%

 




\section{Discrete Zak Transform}\label{sec:DZT}
\input{DZT}

\section{OTFS via OFDM}\label{sec:overlay}
\input{overlay}

\section{OTFS via DZT}\label{sec:OTFS_Zak}
\input{OTFS}

\section{Delay Doppler Spreads}\label{sec:DD}
\input{DD}

\section{Conclusion}\label{sec:conclusion}
\input{Conclusion}


\appendices
\section{Proof of Relation \eqref{eq:Janssen}}
\label{app:proof_DFT_Zak}
\input{Appendices/appendix_DFT_Zak}
\section{Proof of Relation \eqref{eq:IDZTfreq}}
\label{app:proof_inv}
\input{Appendices/appendix_inv_DZT}
\section{Proof of Relation \eqref{eq:DZTproductInv}}
\label{app:proof_prod}
\input{Appendices/appendix_prod} 
\section{Proof of the Modulation Property}
\label{app:proof_mod}
\input{Appendices/appendix_mod}

\section{Proof of the Convolution Property}
\input{Appendices/appendix_conv}

\label{app:proof_conv}


%



\section*{Acknowledgment}
\input{acknowledgment}


\ifCLASSOPTIONcaptionsoff
  \newpage
\fi



%



\bibliography{IEEEabrv,bibliography.bib}

%








\end{document}

%% file: Introduction.tex
\IEEEPARstart{M}{otivated} by challenges encountered in wireless communication over time-variant channels such as Doppler dispersion or  equalization, a new modulation technique termed orthogonal time frequency space (OTFS) was introduced in \cite{monk2016otfs}. The driving idea behind OTFS is to utilize the delay-Doppler (DD) domain to represent information-carrying symbols. The interaction of the corresponding OTFS waveform with a time-frequency (TF) dispersive channel results in a two-dimensional convolution of the symbols in the DD domain \cite[Sec.~III-A]{Hadani3}. OTFS thus turns a time-variant channel into a time-invariant interaction in the DD domain. 
The time-invariant waveform-channel interaction is utilized by OTFS and allows to outperform orthogonal frequency division multiplexing (OFDM) in many scenarios, as shown in \cite{monk2016otfs,Hadani1,Hadani2,Hadani3,Raviteja,Gaudio}.

When OTFS was introduced, it was presented as an overlay for OFDM systems (see \cite[Sec.~III-B]{Hadani3}). The OFDM implementation of OTFS uses the so-called symplectic finite Fourier transform (SFFT) and the inverse SFFT (ISFFT). This transform pair allows mapping symbols defined in the DD domain to the TF domain and vice versa. The overlay for OFDM was initially intended to allow for the reuse of existing hardware \cite[Sec.~IV-A]{Hadani3}, but later became the standard formulation for research on OTFS. For instance, the overlay for OFDM was used in \cite{Raviteja,Gaudio,Wei} to study the input-output relation for TF dispersive channels in the DD domain.

The OTFS overlay for OFDM represents one approach to obtain the time domain signal from its DD representation. Like the Fourier transform, which represents the \emph{fundamental} transform associated with OFDM, the so-called Zak transform can be associated with OTFS, as pointed out in \cite{hadani2018otfs} and \cite[Sec.~III-A]{Hadani3}. In this context, the DD domain and the Zak domain are equivalent. Instead of mapping the symbols residing in the DD first to the TF domain and then to the time domain, the Zak transform directly maps the symbols onto a time-domain signal. A formulation of OTFS based on the \emph{continuous} Zak transform was recently presented in \cite{Khan}.

The OTFS overlay for OFDM as presented in \cite{Hadani3, Raviteja, Gaudio}, and the \emph{continuous} Zak transform approach in \cite{Khan} are based on analog representations of the modulation and demodulation. The analog representations offer a framework to study OTFS and its properties. However, they are ill-suited for practical implementation. For instance, the hardware effort for analog OFDM is prohibitive, as each subcarrier would require a local oscillator \cite[Sec.~19.3]{Molisch}. Similarly, the OTFS modulation presented in \cite{Khan} maps each symbol in the DD domain on an orthogonal basis function. The number of symbols defined on a grid in the DD domain is typically in the order of $10^3-10^4$ (see for example \cite[Sec.~VI]{Gaudio}) or \cite[Sec.~VI]{Raviteja}). Thus, a direct implementation of OTFS based on the \emph{continuous} Zak transform exceeds the hardware effort of the analog implementation of OFDM and makes it even less suitable for direct implementations. 

The main contribution of the paper is to propose OTFS based on the \emph{discrete} Zak transform (DZT), which is motivated by the digital implementation of OFDM (see \cite[Sec.~19.3]{Molisch}, \cite[Sec.~6.4.2.]{barry2004digital}, or \cite[Sec.~4.6]{Stuber}). Unlike the work presented in \cite{Khan}, we do not create an OTFS waveform by introducing time and bandwidth limitations of the \emph{continuous} Zak transform. Instead, we use the DZT, which can be obtained by discretizing the Zak domain \cite{Helmut}. Second, based on the DZT, we derive the input-output relation of OTFS for TF dispersive channels. More specifically, by utilizing DZT signal transform properties, we show that the effect of TF dispersive channels corresponds to convolutions in the delay and Doppler domain. This approach not only simplifies the input-output relation but also furthers the understanding of OTFS. 

We organize the remainder of the paper as follows. In Section~\ref{sec:DZT}, we provide a tutorial-like introduction to the DZT. Then, we establish the connection between DZT and the OTFS overlay for OFDM in Sec.~\ref{sec:overlay}. An  implementation of OTFS based on DZT is presented in Sec.~\ref{sec:OTFS_Zak}. We further establish the input-output relation of OTFS based on the DZT in Sec.~\ref{sec:DD}. Conclusions are drawn in Sec.~\ref{sec:conclusion}.

%% file: DZT.tex
The \emph{continuous} Zak transform is a mapping of a continuous-time signal onto a two-dimensional function. Implicit usage of the Zak transform can be traced back until \emph{Gauss} \cite{Schempp}; however, it was \emph{Zak} who formally introduced the transform in \cite{Zak} and after whom it was named. An excellent paper from a signal theoretical point of view was provided by \textit{Janssen} \cite{Janssen1988}. Later on, \textit{Bölcskei and Hlawatsch} \cite{Helmut} provided an overview of the discrete versions of the transform: the discrete-time Zak transform and the \emph{discrete} Zak transform. This section is devoted to the DZT and its properties since we will use it to describe OTFS and to establish the input-output relation of TF dispersive channel in Sec.~\ref{sec:OTFS_Zak}.

\subsection{Definition and Relations}
In the following discussion, we will treat finite-length sequences of length $KL$ as one period of a periodic sequence with period $KL$. Following the notation in \cite{Helmut}, we use $Z_x^{(L,K)} \in \mathbb{C^{Z\times Z}}$ to denote the DZT of a sequence $x\in \mathbb{C}^\mathbb{Z}$ with period $KL$. The DZT of $x$ is then defined as \cite[eq.~(30)]{Helmut}
\begin{equation}
    Z_x^{(L,K)}[n,k] \triangleq\frac{1}{\sqrt{K}}\sum_{l = 0}^{K-1}\underbrace{x[n+lL]}_{\xdown[l]}e^{-j2\pi \frac{k}{K}l},\quad  n,k\in \mathbb{Z}.\label{eq:DZT}
\end{equation}
It follows from \eqref{eq:DZT} that the DZT for a given $n$, is the unitary discrete Fourier transform (DFT) of a subsampled sequence $\xdown\triangleq \{\xdown[l]= x[n+lL]|l\in \mathbb{Z}\}$. The variable $n$ determines the starting phase of the downsampled sequence, whereas the variable $k$ is the discrete frequency of its DFT. Thus, the variables $n$ and $k$ represent time and frequency, respectively.  

The sequence $x$ can be recovered from its DZT through the following sum relation:
\begin{equation}
x[n] = \frac{1}{\sqrt{K}} \sum_{k=0}^{K-1} Z_{x}^{(L,K)}[n,k],
\label{eq:IDZTtime}
\end{equation}
which follows from the definition of the DZT in \eqref{eq:DZT} and the relation 
\begin{equation}
    \sum_{k=0}^{K-1}e^{-j2\pi \frac{l}{K}k}= K\sum_{m=-\infty}^{\infty} \delta[l-mK],
    \label{eq:delta_train}
\end{equation}
where $\delta[n]$ denotes the Kronecker delta. We will refer to \eqref{eq:IDZTtime} as inverse DZT (IDZT).

\emph{Notation:} Depending on the period of the sequence under consideration ($KL$), different choices of $K$ and $L$ are possible. 
We indicate the choice of $K$ and $L$ in the superscript of the DZT notation we use ($Z_x^{(L,K)}$). If the choice is not important for the context, we will drop the superscript for brevity of notation $(\Zxs)$. Furthermore, the DZT is in general a complex-valued function. To illustrate the DZT, we will often write the DZT in polar form, i.e.,
\begin{equation}
    \Zxs[n,k]=\left|\Zxs[n,k]\right|e^{j\varphi_x [n,k]},
\end{equation}
where $\left|\Zxs[n,k]\right|$ and $\varphi_x [n,k]$ represent the magnitude and the phase of $\Zx$, respectively. We restrict the phase to the principal values, i.e., to the interval $[-\pi,\pi)$.

\begin{myExample}[DZT]
\label{ex:DZT}
Consider the $KL$-periodic sequence $g$ with elements
\begin{equation}
    g[n] = 
    \begin{cases} f[n], & 0\leq n \leq L-1, \\
                0, & L \leq n \leq KL-1.
                \label{eq:piecewise}
    \end{cases}
\end{equation}
The sequence is zero almost everywhere, except for the first $L$ samples where it takes the value of an arbitrary sequence $f$. The second condition in \eqref{eq:piecewise} implies that only one nonzero addend (for $l=0$) exists in the summation \eqref{eq:DZT}. Thus, the elements of $Z_g$ are 
 \begin{equation}
 \Zgs = e^{j2\pi\frac{k}{K}\lfloor n/L \rfloor} \frac{1}{\sqrt{K}} f[(n)_{L}].
 \label{eq:DZT_ex}
 \end{equation}
Here, $(\cdot)_L$ denotes the modulo $L$ operation and $\lfloor n/L \rfloor$ is the greatest integer less than or equal to $n/L$. An example for a sequence $f$ and the corresponding magnitude of the DZT $Z_g$, are illustrated in Fig.~\ref{fig:DZTex}a) and b), respectively. 
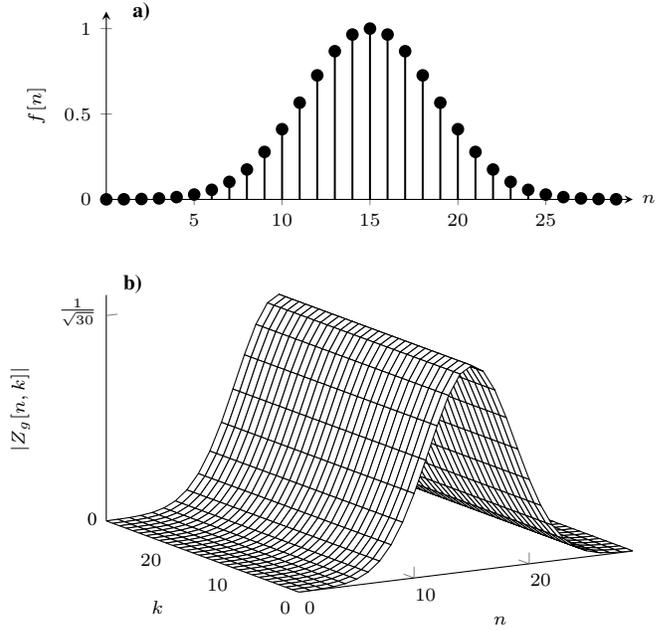
\begin{figure}
    \centering
    \input{Figures/g_Zg}
    \caption{a) Sequence $f[n]=e^{-\frac{1}{2}\left(\frac{n-L/2}{\sigma L/2}\right)^2}$ for $\sigma = 1/4$, $0\leq n \leq L-1$ and $L=30$. The sequence $g$ has period $KL=900$. b) Magnitude of the DZT $Z_g$ with parameters $K=30$, $L=30$ in \eqref{eq:DZT_ex}, for $0\leq n\leq L-1$ and $0\leq k \leq K-1$. The phase $\varphi_g[n,k]$ (not plotted) is zero for the presented values of $n$ and $k$, see \eqref{eq:DZT_ex}.}
    \label{fig:DZTex}
\end{figure}
\end{myExample}

We express the period of the sequence $x$ as a product $KL$ with $K, L\in\mathbb{N}$. This factorization ensures that the sequence can be decomposed into $L$ subsampled sequences with period $K$. In general, the product $KL$ is not unambiguously defined since different choices of $K$ and $L$ will result in the same product. Independent of the period, two choices are always possible and will provide some interesting insights. Firstly, the choice $K=1$ in \eqref{eq:DZT} leads to
\begin{equation}
        Z_{x}^{(L,1)}[n,k] =  x[n],
        \label{eq:dzt_K1}
\end{equation}
i.e., the elements of DZT for a specific $n$ and any $k$ are the elements of the sequence $x$.  Secondly, the case $L = 1$ results in  
\begin{equation}
    Z_{x}^{(1,K)}[n,k] = \frac{1}{\sqrt{K}}\sum_{l = 0}^{K-1}x[n+l]e^{-j2\pi \frac{k}{K}l}.
    \label{eq:dzt_dft}
\end{equation}
For $n=0$, we obtain 
\begin{equation}
      Z_{x}^{(1,K)}[0,k] = X[k]
      \label{eq:DFT}
\end{equation}
where $X\in \mathbb{C^Z}$ is the unitary DFT of the sequence $x$, i.e.,
\begin{equation}
    X[k] \triangleq \frac{1}{\sqrt{K}}\sum_{l = 0}^{K-1}x[l]e^{-j2\pi \frac{k}{K}l}.
    \label{eq:DFT_1}
\end{equation}
It follows from \eqref{eq:dzt_dft} that $Z_x^{(1,K)}[n,k]$ represents the DFT of the circular shifted sequence $x$ with shift parameter $n$.
Using the circular shift property of the DFT given as \cite[eq.~(3.168)]{Vetterli})
\begin{equation}
    x[n-n_0]  \Leftrightarrow e^{-j2\pi\frac{k}{K}n_0} X[k],
    \label{eq:DFT_shift}
\end{equation}
we can express \eqref{eq:dzt_dft} equivalently as
\begin{equation}
        Z_{x}^{(1,K)}[n,k]  =  e^{j2\pi\frac{k}{K}n}X[k]=  e^{j2\pi\frac{k}{K}n}Z_{x}^{(1,K)}[0,k].
        \label{eq:DFT_shifted}
\end{equation}

Following the same approach that provided the DFT \eqref{eq:DFT}, we can obtain the inverse DFT (IDFT). Therefore, we consider \eqref{eq:IDZTtime} for the case $L=1$, which is
\begin{equation}
    x[n] \triangleq \frac{1}{\sqrt{K}}\sum_{k=0}^{K-1} X[k]e^{j2\pi\frac{k}{K}n},
    \label{eq:IDFT}
\end{equation}
where \eqref{eq:IDFT} is obtained from substituting  \eqref{eq:DFT_shifted} in \eqref{eq:IDZTtime}.

The DZT $Z_x$ of a sequence $x$ cannot only be obtained from a sequence $x$, but also from its DFT $X$ in \eqref{eq:DFT} through
\begin{equation}
        \Zx=\frac{1}{\sqrt{L}}\sum_{l=0}^{L-1}X[k+lK]e^{j2\pi\frac{k+lK}{KL}n}.
    \label{eq:Janssen}
\end{equation}
\begin{proof}
See Appendix~\ref{app:proof_DFT_Zak}.
\end{proof}

Equivalently, using \eqref{eq:DZT}, we recognize \eqref{eq:Janssen} as
\begin{equation}
     \Zx = e^{j2\pi\frac{n}{KL}k}Z_X^{(K,L)}[k,-n] \label{eq:Janssen2},
\end{equation}
where $Z_X^{(K,L)}$ is the DZT of the DFT sequence $X$.

The corresponding inverse relation is
\begin{equation}
X[k] = \frac{1}{\sqrt{L}} \sum_{n=0}^{L-1}\Zx e^{-2\pi\frac{k}{KL}n}.
\label{eq:IDZTfreq}
\end{equation}
\begin{proof}
See Appendix~\ref{app:proof_inv}.
\end{proof}

Fig.~\ref{fig:relations} summarizes the relations between the sequence $x$, the DZT $\Zxs$, and the DFT $X$. Note that the DFT $X$ can be obtained in two ways. Either directly via \eqref{eq:DFT_1} or indirectly using \eqref{eq:DZT} and \eqref{eq:IDZTfreq}. The later approach resembles the Cooley-Tuckey algorithm, a fast Fourier transform algorithm \cite{Helmut}. 

\begin{figure}
    \centering
    \input{Figures/relations}
    \caption{Different signal representations of a sequence $x$ and its corresponding transforms: DZT $\Zxs$ and DFT $X$.} 
    \label{fig:relations}
\end{figure}
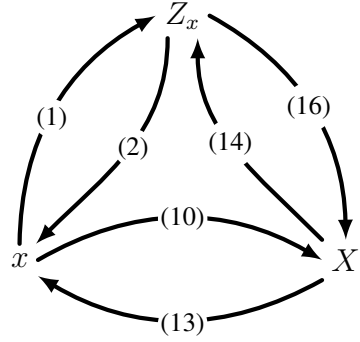

\subsection{Properties of the DZT}
The DFT $X$ of a sequence $x$ with length $K$ is periodic with period $K$, i.e., $X[k] = X[k+mK]$ with $m\in \mathbb{Z}$, see \eqref{eq:DFT_1}. The DZT possess similar properties since the DZT is the DFT of the downsampled sequence $\xdown$, see \eqref{eq:DZT}. Consequently, the DZT is also periodic in the frequency variable $k$, i.e.,
\begin{equation}
    Z_{x}^{(L,K)}[n,k+mK] =  Z_{x}^{(L,K)}[n,k], \quad m \in \mathbb{Z}.
    \label{eq:periodic}
\end{equation}
By the circular shift property of the DFT in \eqref{eq:DFT_shift}, we further have
\begin{equation}
    Z_{x}^{(L,K)}[n+mL,k] = e^{j2\pi \frac{k}{K}m}  Z_{x}^{(L,K)}[n,k],\quad m \in \mathbb{Z} ,
    \label{eq:quasi_periodic}
\end{equation}
i.e., the DZT is periodic in $n$ with period $L$ up to a complex factor $e^{j2\pi(k/K)m}$. The DZT is therefore said to be \emph{quasi}-periodic with \emph{quasi}-period $L$. Due to the periodicity properties in \eqref{eq:periodic} and \eqref{eq:quasi_periodic}, the DZT is fully determined by the DZT for $0\leq n \leq L-1$ and $0\leq k \leq K-1$, which is referred to as the fundamental rectangle \cite{Helmut}.

The \emph{quasi}-periodicity in \eqref{eq:quasi_periodic} can be utilized to express the IDZT in \eqref{eq:IDZTtime} as 
\begin{equation}
    x[n+lL] = \frac{1}{\sqrt{K}} \sum_{k=0}^{K-1} \Zx e^{j2\pi\frac{k}{K}l}.
    \label{eq:IDZT2}
\end{equation}
Here, we express the index of the sequence as sum of the form $n+lL$ with $0\leq n \leq L-1$ and $l \in \mathbb{Z}$. Since the fundamental rectangle fully determines the DZT $Z_x$, we will restrict ourselves to this fundamental rectangle when plotting the DZT. In fact, this is what we already did in Fig.~\ref{fig:DZTex}b).

\begin{myExample}[IDZT]
\label{ex:IDZT}
Consider the DZT that is defined by a single nonzero coefficient on the fundamental rectangle of size $4\times 6$ and is given by
\begin{equation}
Z_{x}^{(4,6)}[n,k] = \delta[n]\delta[k].
\label{eq:exIDZT1}
\end{equation}
The fundamental rectangle and the DZT in \eqref{eq:exIDZT1} are illustrated in Fig.~\ref{fig:impuls_train}a) (left). One period of the sequence $x$ obtained through \eqref{eq:IDZT2} is 
\begin{equation}
    x[n] = \frac{1}{\sqrt{6}} \sum_{l=0} ^{K-1} \delta[n-6l],
    \label{eq:IDZT_seq1}
\end{equation}
i.e., a train of real Kronecker deltas starting at $n=0$ with spacing $L=6$, as shown in Fig.~\ref{fig:impuls_train}a) (right). 
Consider now the DZT
\begin{equation}
Z_{y}^{(4,6)}[n,k] =\delta[n-3]\delta[k-5],
\label{eq:exIDZT2}
\end{equation}
which is shown in Fig.~\ref{fig:impuls_train}b). One period of the corresponding sequence $y$ is
\begin{equation}
    y[n] = \frac{1}{\sqrt{6}} \sum_{l=0}^{K-1}\delta[n-3-6l] e^{j2\pi\frac{5}{6}l}
\end{equation}
and is shown in Fig.~\ref{fig:impuls_train}b). When compared to $x$, the sequence $y$ is delayed by three samples and modulated with a discrete frequency $k=5$.

In fact, a single coefficient at $Z_x[n,k]$ maps onto a sequence 
\begin{equation}
    v_{n,k}[n'] =
    \frac{1}{\sqrt{K}}\sum_{l=0}^{K-1} \delta[n'-n+lL]e^{j2\pi\frac{k}{K}l}.
\end{equation}
The set of sequence $\{v_{n,k}:0\leq n\leq L-1,0\leq k\leq K-1\}$ forms an orthonormal basis and $Z[n,k]$ are the \emph{expansion coefficients}. We will use this fact in Sec.~\ref{sec:OTFS_Zak}, where we define a sequence by its corresponding DZT in the same way as OFDM defines the symbols in the DFT domain.

\begin{figure}
    \centering
    \input{Figures/impulse_trains}
    \caption{Two examples of DZTs (left) defined by a single nonzero coefficient on the fundamental rectangle (indicated by a dot) and the corresponding sequences (right) for a) the DZT given in  \eqref{eq:exIDZT1},  and b) given in \eqref{eq:exIDZT2}.}
    \label{fig:impuls_train}
\end{figure}
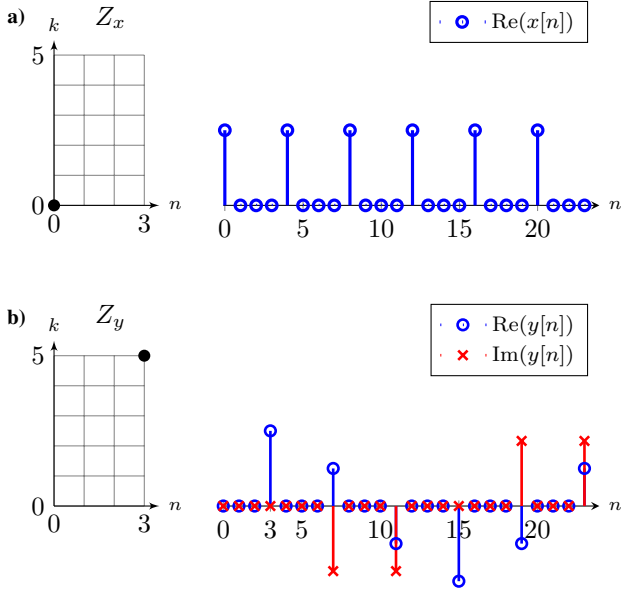
\end{myExample}

By the \emph{quasi}-periodicity we further have that the elementwise product of a DZT $Z_x$ with the complex conjugate DZT $Z_y^{\ast}$ is periodic in $n$ and $k$. Motivated by the periodicity, we can apply a two-dimensional DFT, which turns out to be \cite{Helmut, An1998}
\begin{equation}
     \sum_{n=0}^{L-1}\sum_{k=0}^{K-1} \Zxs[n,k]  Z_{y}^\ast[n,k] e^{j2\pi\left(\frac{m}{K}k-\frac{l}{L}n\right)} = \langle x, y_{m,l} \rangle,
    \label{eq:DZTproductInv}
\end{equation}
where $y_{m,l}\triangleq y[n-mL]e^{j2\pi(l/L)n}$. Note that the Fourier kernel in \eqref{eq:DZTproductInv} has opposed signs for the two individual dimensions.
\begin{proof}
See Appendix~\ref{app:proof_prod}.
\end{proof}
The inverse relation is given by
\begin{equation}
    \Zxs[n,k] Z_{y}^\ast[n,k] = \frac{1}{KL} \sum_{m=0}^{K-1}\sum_{l=0}^{L-1} \langle x, y_{m,l} \rangle e^{-j2\pi\left(\frac{k}{K}m-\frac{n}{L}l\right)},
    \label{eq:DZTproduct}
\end{equation}
which follows from applying the corresponding two-dimensional inverse DFT on both sides of \eqref{eq:DZTproductInv}. The relations \eqref{eq:DZTproductInv} and \eqref{eq:DZTproduct}  provide a useful tool when considering the OTFS overlay for OFDM in Sec.~\ref{sec:overlay}.

\subsection{Signal Transform Properties}\label{subsec:STP}
Here, we will list three signal transform properties that we will use later in the study of OTFS. A comprehensive overview of signal transform properties can be found in \cite[Table~VII]{Helmut}. Let $x$, $y$, and $z$ are sequences with same periods and let $Z_x$, $Z_y$, and $Z_z$ be there DZTs. Then following properties hold:
\begin{enumerate}
    \item \textit{Shift:}
    Let $y$ be the shifted version of $x$, i.e., $y[n]=x[n-m]$, then
    \begin{equation}
        \Zys[n,k] = Z_x[n-m,k].
        \label{eq:dzt_shift}
    \end{equation}
     A shift of the sequence causes a shift of the corresponding DZT. The proof follows from the definition of the DZT \eqref{eq:DZT}. For shifts of multiples of $L$, i.e., $m=lL$ with $l\in \mathbb{Z}$, we further have 
     \begin{equation}
         \Zys[n,k] = e^{-j2\pi\frac{k}{K}m}Z_x[n,k],  
         \label{eq:shiftL}
     \end{equation}
     which follows from the \emph{quasi}-periodicity of the DZT in 
     \eqref{eq:quasi_periodic}.
   
    \item \textit{Modulation:}
     Let $z=x\cdot y$ be the elementwise product of $x$ and $y$, i.e., ${z[n]=x[n] y[n]}$. Then,
    \begin{equation}
        Z_z[n,k] = \frac{1}{\sqrt{K}}\sum_{l=0}^{K-1} Z_x[n,l]Z_y[n,k-l],
        \label{eq:modProp}
    \end{equation}
    i.e., the DZT of the element-wise multiplication is a scaled convolution with respect to the variable $k$.
    \begin{proof}
    See Appendix~\ref{app:proof_mod}.
    \end{proof}
    
    \item \textit{Circular Convolution:}
    Consider $z=x\circledast y$, i.e., the circular convolution of $x$ and $y$. The DZT $Z_z$ is
    \begin{equation}
        Z_z[n,k] = \sqrt{K}\sum_{m=0}^{L-1} Z_x[m,k]Z_y[n-m,k],
        \label{eq:convProp}
    \end{equation}
    i.e., the DZT of a circular convolution is the scaled convolution with respect to the variable $n$ up to a constant.
    \begin{proof}
    See Appendix~\ref{app:proof_conv}.
    \end{proof}
\end{enumerate}

\begin{figure}
    \centering
    \includegraphics[]{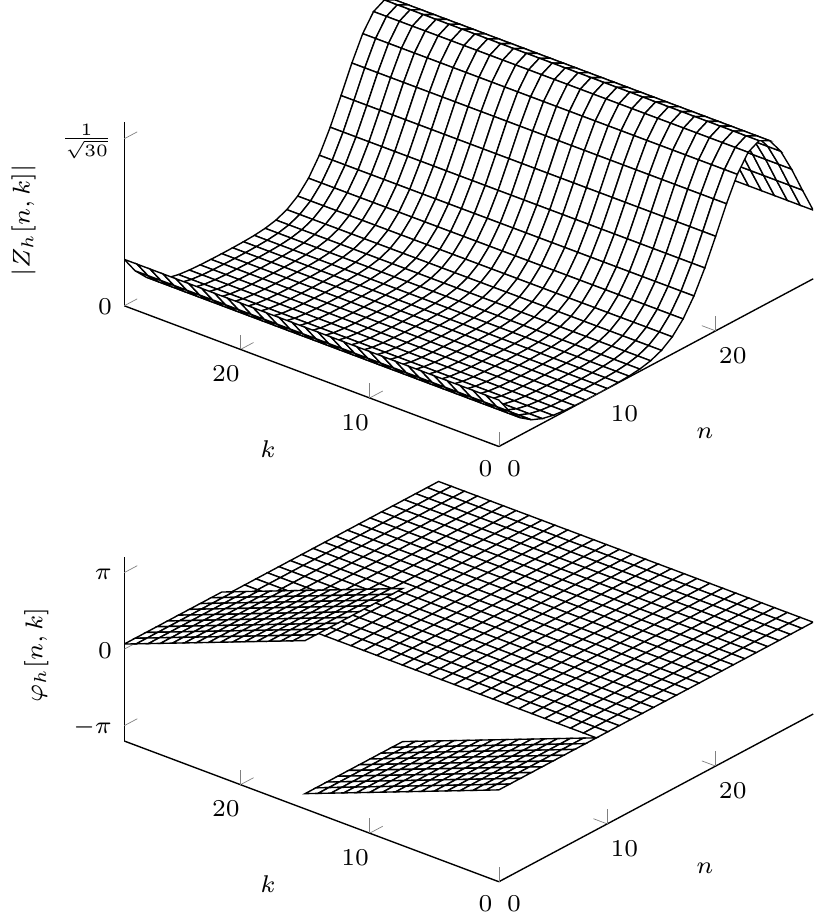}
    \caption{The DZT $Z_h[n,k] = Z_{g}[n-10,k]$ in Example~\ref{ex:shift}, with $\Zgs$ being the DZT of Fig.~\ref{fig:DZTex}. The shift of the DZT with respect to $n$ causes a circular shift of the magnitude $|Z_g[n,k]|$ of DZT (top). The phase $\varphi_h[n,k]$ experiences an additional linear phase for indices smaller than 10 (bottom).} 
    \label{fig:DZTshift}
\end{figure}

The shift property in \eqref{eq:dzt_shift} together with the \emph{quasi}-periodicity in \eqref{eq:quasi_periodic} has also another important implication. In OTFS, as we will show in Sec.~\ref{sec:OTFS_Zak}, the received signal will include a superposition of delayed sequences that are in general not multiples of $L$. We will discuss this in Example~\ref{ex:shift}. 

\begin{myExample}[Shifted DZT]\label{ex:shift}
Consider the DZT $Z_h$ with elements
\begin{equation}
    Z_{h}[n,k] = Z_{g}[n-10,k],
\end{equation}
which is a shifted version of the DZT $Z_g$ in Fig.~\ref{fig:DZTex}.b) of Example~\ref{ex:DZT}. To evaluate the DZT $Z_h$ within the fundamental rectangle, we first make the observation that any index $n$ can be expressed as $n = i+mL$ with $m =\lfloor n/L  \rfloor$.
In this example, the indices $n=0$ to $9$ of $Z_h$ correspond to the indices $n=-10$ to $-1$ of $Z_g$. Expressing the latter indices in terms of $i$ and $m$, we have $m=-1$ and $i$ from 20 to 29. Thus, by the \emph{quasi}-periodicity property in \eqref{eq:quasi_periodic}, we have that the magnitude of  DZT $Z_h[n,k]$  is   $|Z_h[n,k]| = |Z_g[n+20,k]|$ for $20\leq n \leq 29$. However, the phase $\varphi_h[n,k]$ exhibits an additional linear phase term and is given as $e^{-j2\pi k/K} \varphi_h[n+20,k]$. On the other hand, the indices of $10 \leq n \leq 29$ of $Z_{h}[n,k]$ correspond to the indices $0 \leq n \leq 19$ of $Z_g[n,k]$. Therefore, $m=0$ and $Z_h$ is the shifted DZT $Z_g$ within the fundamental rectangle. Thus,
\begin{equation}
    Z_h[n,k] = \begin{cases}
                e^{-j2\pi \frac{k}{K}}Z_{g}[20+n,k], & 0\leq n \leq 9, \\
                Z_{g}[n-10,k], & 10\leq n \leq 29,
    \end{cases}
\end{equation}
or more generally, $Z_h[n,k] =e^{j2\pi (k/K)\lfloor (n-10)/L\rfloor}{Z_{g}[(n-10)_L,k]}$. The DZT $Z_h$ is depicted in Fig.~\ref{fig:DZTshift}, which illustrates also different phase behaviors.

\end{myExample}

%% file: Figures/g_Zg.tex
\begin{tikzpicture}
\begin{axis}[ 
    name = main plot,
    width = 7cm,
    height = 2.5cm,
    grid=none,
    scale only axis = true,
    xmin=0,xmax=29.95,
    ymin=0,ymax=1.1,
    axis x line = middle,
    axis y line = left,
    xlabel = {$n$},
    ylabel = {$f[n]$},
    x label style={anchor=west},
    label style={font=\scriptsize},
    tick label style={font=\scriptsize}  
  ]
\addplot[ycomb,mark=*, color = black, domain=0:29, samples = 30, line width=0.75pt] {exp(-1/2*((x-15)/(1/4*15))^2)};
\node[](a) at (axis cs: 2,1.1) {};
\end{axis}
\node[] at (a) {\footnotesize\textbf{a)}};

\path (main plot.below south west) ++(0,-4.75cm)
coordinate (lower plot position);

\begin{axis}[
    height = 4.5cm,
    width = 7cm,
    at = {(lower plot position)},
    view={-30}{20},
    grid=none,
    xmin=0,xmax=29,
    ymin=0,ymax=29,
    zmin=0,
    axis lines*=left,
    scale only axis = true,
    trig format plots=rad,
    xlabel = {$n$},
    ylabel = {$k$},
    zlabel = {$|\Zgs|$},
    every axis z label/.style={at={(rel axis cs:-0.2,1.1,0.5)},rotate = 90},
    ztick = {0,1},
    zticklabels={0,{$\frac{1}{\sqrt{30}}$}},
    label style={font=\scriptsize},
    tick label style={font=\scriptsize},
  ]
  \addplot3[surf,faceted color = black, domain=0:29, samples = 30,y domain=0:29, samples y = 30,colormap={bw}{gray(0cm)=(1); gray(1cm)=(1)}] {exp(-1/2*((x-15)/(1/4*15))^2)};

\node[](b) at (axis cs: 0,25,1.2) {};
\end{axis}
\node[] at (b) {\footnotesize\textbf{b)}};

\end{tikzpicture}

%% file: Figures/relations.tex
\usetikzlibrary{decorations.pathmorphing,arrows,calc}

\begin{tikzpicture}[line cap=round, line join=round]
\node[] (x) at (210:2.5) {\large $x$};
\node[] (X) at (-30:2.5) {\large $X$};
\node[] (Zx) at (90:2) {\large $Z_{x}$};

\draw [ ultra thick, -latex] (x) to [bend left](Zx.west);
\draw [ ultra thick, latex-] (x) to [out = 45, in = 270]  ($(Zx.south)-(0.2,0)$);
\draw [ ultra thick, latex-](X) to [bend right] (Zx.east);
\draw [ ultra thick, -latex] (X.north west) to [out= 135, in = 270] ($(Zx.south)+(0.2,0)$);
\draw [ ultra thick, latex-] (x.south east) to [bend right](X.south west);
\draw [ ultra thick, -latex](x.east) to [bend left](X.west);

\node[fill = white, circle,inner sep=0pt] (rel1) at (160:1.85) {\eqref{eq:DZT}};
\node[fill = white, circle,inner sep=0pt] (rel2) at (160:0.7) {\eqref{eq:IDZTtime}};
\node[fill = white, circle,inner sep=0pt] (rel3) at (25:1.85) {\eqref{eq:IDZTfreq}};
\node[fill = white, circle,inner sep=0pt] (rel4) at (25:0.7) {\eqref{eq:Janssen}};
\node[fill = white, circle,inner sep=0pt] (rel5) at (270:2.1) {\eqref{eq:IDFT}};
\node[fill = white, circle,inner sep=0pt] (rel6) at (270:0.7) {\eqref{eq:DFT_1}};
\end{tikzpicture}

%% file: Figures/impulse_trains.tex
\begin{tikzpicture}

\begin{scope}[scale = 0.4]
\draw[gray,very thin] (0,0) grid (3,5);
\draw[-latex'] (0,0) -- (3.5,0)node[anchor = west]{\scriptsize $n$};
\draw[-latex'] (0,0) -- (0,5.5)node[anchor = south]{\scriptsize $k$};
\node[anchor = north] at (0,0) {$0$};
\node[anchor = north] at (3,0) {$3$};
\node[anchor = east] at (0,0) {$0$};
\node[anchor = east] at (0,5) {$5$};
\fill[black] (0,0) circle (0.2cm);
\end{scope}

\begin{axis}[
at = {(2.25cm,0)},
name = main plot,
width = 5cm,
height = 3cm,
scale only axis = true,
y=1cm,
axis x line=center,
axis y line=center,
hide obscured x ticks=false,
xmax=23.95,
xmin=-0.1,
ytick = \empty,
y axis line style={draw=none},
xlabel={$n$},
x label style={anchor=west, font = \scriptsize},
legend style={legend pos = north east, yshift = 1.75cm},
]
\addplot[ycomb, blue,mark=o, domain = 0:23, samples = 24, line width =1.2pt]{mod(x,4)==0};
\addlegendentry{\footnotesize $\operatorname{Re}(x[n])$}
\end{axis}
\node[] at (-0.5,2.5) {\footnotesize\textbf{a)}};
\node[] at (0.75,2.5) {$Z_x$};

\begin{scope}[shift = {(0,-4cm)}, scale = 0.4]
\draw[gray,very thin] (0,0) grid (3,5);
\draw[-latex'] (0,0) -- (3.5,0)node[anchor = west]{\scriptsize $n$};
\draw[-latex'] (0,0) -- (0,5.5)node[anchor = south]{\scriptsize $k$};
\node[anchor = north] at (0,0) {$0$};
\node[anchor = north] at (3,0) {$3$};
\node[anchor = east] at (0,0) {$0$};
\node[anchor = east] at (0,5) {$5$};
\fill[black] (3,5) circle (0.2cm);
\end{scope}

\node[] at (-0.5,-1.5) {\footnotesize\textbf{b)}};
\node[] at (0.75,-1.5) {$Z_y$};

\begin{axis}[
at = {((2.25cm,-4cm)},
anchor = west,
width = 5cm,
scale only axis = true,
y = 1cm,
axis x line=center,
axis y line=center,
extra x ticks={3},
scale only axis = true,
hide obscured x ticks=false,
trig format plots=rad,
xmax=23.95,
ytick = \empty,
y axis line style={draw=none},
xlabel={$n$},
x label style={anchor=west, font = \scriptsize},
legend style={legend pos = north east, yshift = 1.75cm},
]
\addplot[ycomb, blue,mark=o, domain = 0:23, samples = 24,line width =1pt]{(mod(x-3,4)==0)*cos(2*pi*5/6*int(x/4))};
\addlegendentry{\footnotesize $\operatorname{Re}(y[n])$}; 
\addplot[ycomb, line width =1pt, red,mark=x,mark options ={scale=1.25},red, domain = 0:23, samples = 24]{(mod(x-3,4)==0)*sin(2*pi*5/6*int(x/4))};
\addlegendentry{\footnotesize $\operatorname{Im}(y[n])$}; 
\end{axis}

\end{tikzpicture}

%% file: overlay.tex
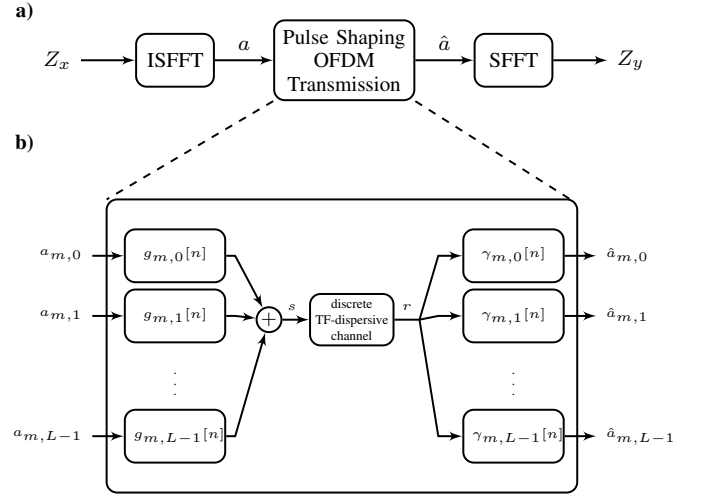
\begin{figure}
    \centering
    \input{Figures/overlay}
    \caption{a) Conceptual description of OTFS as an overlay technique: the ISFFT and SFFT are used to transform the symbols from the DD domain to the TF domain and vice versa. The symbols in the TF domain are used as an input for a pulse shaping OFDM transmission shown in b).}
    \label{fig:overlay}
\end{figure}

In this section, we revisit the OFDM-based implementation of OTFS usually used in the literature. In the OFDM implementation, OTFS is considered as an additional pre- and post-processing, given by the ISFFT and SFFT. OTFS literature typically considers pulse shaping OFDM (see for example \cite{Hadani1,Raviteja,Gaudio}), i.e., transmit and receive pulses that are not necessarily rectangular. We consider the discrete version of pulse shaping OFDM \cite[Sec.~3]{Helmut2}. In this discrete setting, we consider all sequences, including the transmit and receive pulses, to be $KL$-periodic. Based on the discrete formulation, we will show that the overlay technique resembles the IDZT/DZT. We further show that the transmitted symbols can be recovered if the transmit and receive pulses satisfy the so-called biorthogonality condition.

The OFDM overlay schematically illustrated in Fig.~\ref{fig:overlay}a) can be described as follows. A block of $L\times K$ complex symbols, typically referred to as a frame, is defined in the DD domain. We consider the DD and the Zak domain equivalent, and thus, we use the notation introduced in Sec.~\ref{sec:DZT} and denote the frame by $\Zxs$. A frame represents the fundamental rectangle of a $KL$-periodic sequence. The frame $\Zxs$ in the DD domain is then mapped to a frame $a$ in the TF domain. The individual symbols $a_{m,l}$ are obtained via the SFFT, which is defined as \cite[eq.~(2)]{Raviteja}
\begin{equation}
    a_{m,l}\triangleq \frac{1}{\sqrt{KL}} \sum_{n=0}^{L-1} \sum_{k=0}^{K-1} \Zxs[n,k]e^{j2\pi\left(\frac{m}{K}k-\frac{l}{L}n\right)},
    \label{eq:ISFFT}
\end{equation}
with $0\leq m \leq K-1$ and  $0\leq l \leq L-1$. 

The $K\times L$ symbols $a_{m,l}$ in the TF domain are the input for pulse shaping OFDM modulation. Here, $L$ is the number of subcarriers, and $K$ is the number of symbols (see Fig.~\ref{fig:overlay}b)). The transmitted signal $s$ is then\footnote{In the OTFS literature, this mapping is sometimes referred to as Heisenberg transform.}
\begin{equation}
    s[n] = \sum_{m=0}^{K-1} \sum_{l=0}^{L-1} a_{m,l} g_{m,l}[n],
    \label{eq:OFDM_mod}
\end{equation}
where ${\gml[n] \triangleq g[n-mL]e^{j2\pi (l/L)n}}$ are TF shifted replicas of the discrete-time transmit pulse $g$, see Fig.~\ref{fig:overlay}b).
The transmitted signal is sent over a TF dispersive channel. The digital to analog conversion and the effect of the channel will be addressed in Sec.~\ref{sec:OTFS_Zak} and \ref{sec:DD}. To obtain the received symbols $\hataml$ in the TF domain, the receiver calculates the inner product between the received signal $r$ and the pulses $\gammaml$, i.e.,
\begin{equation}
   \hataml = \langle r,\gammaml\rangle,
   \label{eq:ofdm_analysis}
\end{equation}
where ${\gammaml[n] \triangleq \gamma[n-mL]e^{j2\pi (l/L)n}}$.

In a final step, the received symbols $\hataml$, residing in the TF domain, are mapped back to the DD domain using the SFFT, defined as \cite[eq.~(11)]{Raviteja}
\begin{equation}
    \Zys[n,k] \triangleq \frac{1}{\sqrt{KL}} \sum_{l=0}^{L-1} \sum_{m=0}^{K-1} \hataml e^{-j2\pi\left(\frac{k}{K}m-\frac{n}{L}l\right)},
    \label{eq:SFFT}
\end{equation}
where $Z_y$ denotes the received block in the DD domain.

In pulse shaping OFDM, we have, in the absence of the channel ($r=s$), that $\aml = \hataml$ if the pulse pair $g$ and $\gamma$ satisfies the so-called biorthogonality condition given as \cite[Sec.~4]{WEXLER} 
\begin{equation}
    \langle g, \gammaml\rangle = \delta[m]\delta[l]
    \label{eq:biorthogonality}.
\end{equation}

\subsection{Overlay with Rectangular Pulses}
In the particular case of OFDM with rectangular pulses, we have that the transmit and the receive pulses are
\begin{equation}
    g[n]=\gamma[n]=\begin{cases}
    \frac{1}{\sqrt{L}}, & 0\leq n \leq L-1 \\
    0 & L \leq n \leq KL-1,
    \end{cases}
\end{equation}
and \eqref{eq:OFDM_mod} can be expressed as
\begin{align}
    s[n+mL] & = \frac{1}{\sqrt{L}} \sum_{m=0}^{K-1} \sum_{l=0}^{L-1} a_{m,l} e^{j2\pi \frac{k}{L}n}\label{eq:OFDM_rect}\\
    & =  \frac{1}{\sqrt{K}} \sum_{k=0}^{K-1}\Zxs[n,k]e^{j2\pi\frac{m}{K}k}\label{eq:OFDM_IDZT}
\end{align}
which follows from substituting \eqref{eq:ISFFT} in \eqref{eq:OFDM_rect} and using \eqref{eq:delta_train}. We recognize \eqref{eq:OFDM_IDZT} as the IDZT in \eqref{eq:IDZT2}. Similarly, using \eqref{eq:DZTproductInv} and the fact that the DZT of the rectangular receive pulse $\gamma$ is 
\begin{equation}
    Z_{\gamma}[n,k] = \frac{1}{\sqrt{KL}},
\end{equation}
 we get
\begin{equation}
 \frac{1}{\sqrt{KL}} Z_y[n,k] =\frac{1}{KL} \sum_{m=0}^{K-1}\sum_{l=0}^{L-1} \langle y, \gammaml \rangle e^{-j2\pi\left(\frac{k}{K}m-\frac{n}{L}l\right)}.
\label{eq:SFFT_derivation}
\end{equation}
Multiplying both sides of \eqref{eq:SFFT_derivation} by $\sqrt{KL}$, we obtain
\begin{equation}
 Z_y[n,k] =\frac{1}{\sqrt{KL}} \sum_{m=0}^{K-1}\sum_{l=0}^{L-1} \langle y, \gammaml \rangle e^{-j2\pi\left(\frac{k}{K}m-\frac{n}{L}l\right)},
\end{equation}
where the left hand side is the DZT of $y$ and the right hand side is the composition of OFDM demodulation in \eqref{eq:DZTproductInv} and the SFFT in \eqref{eq:SFFT}. Thus, in case of rectangular pulse shapes, the overlay technique is equivalent to the DZT. Moreover, the direct implementation using the DZT is beneficial in terms of computational complexity. To see this, consider first the IDZT implementation. Computing the sequence $s$ from the DZT $Z_x$ in \eqref{eq:OFDM_IDZT} requires the computation of $L$ DFTs of length $K$. On the other hand, the overlay requires $K$ DFTs of length $L$ and $L$ IDFTs of length $K$ for the ISFFT \eqref{eq:ISFFT}. Additionally, the OFDM modulation in \eqref{eq:OFDM_mod} requires $K$ IDFTs or length $L$. In fact, the $K$ IDFTs of the OFDM modulation negate the $K$ DFTs of the ISFFT, leaving effectively only the $L$ IDFs of length $K$. Thus, the DZT implementation requires less computation.

\subsection{Overlay with Nonrectangular Pulses}
For nonrectangular pulses $g$ and $\gamma$ we first recognize that after substituting \eqref{eq:ISFFT} in \eqref{eq:OFDM_mod} and rearranging terms we have
\begin{multline}
    s[n] = \frac{1}{\sqrt{L}}  \sum_{k=0}^{K-1} \sum_{n'=0}^{L-1} \sum_{l=0}^{L-1} Z_x[n',k]e^{-2\pi\frac{n'-n}{L}l}\\ \frac{1}{\sqrt{K}}\sum_{m=0}^{K-1} g[n-mL]e^{2\pi\frac{k}{K}m}.
    \label{eq:OFDM_DZT}
\end{multline}
Considering the periodicity of $g$, we recognize the last sum in \eqref{eq:OFDM_DZT} as the DZT of $g$. Using \eqref{eq:delta_train}, we further get that 
\begin{equation}
    s[n] = \frac{1}{\sqrt{K}}\sum_{k=0}^{K-1} \sqrt{KL}\Zxs[n,k] Z_g[n,k],
\end{equation}
which is the IDZT \eqref{ex:IDZT} of the scaled product $\sqrt{KL}\Zxs[n,k] Z_g[n,k]$. Thus, we can define the scaled product as the DZT of $s$, i.e.,
\begin{equation}
     \sqrt{KL}\Zxs[n,k] Z_g[n,k] = Z_s[n,k].
    \label{eq:Zs}
\end{equation}

In the absence of a channel ($r=s$), the composition of OFDM demodulation \eqref{eq:ofdm_analysis} and SFFT \eqref{eq:SFFT} is equivalent to
\begin{multline}
    \frac{1}{\sqrt{KL}} \sum_{m=0}^{K-1}\sum_{l=0}^{L-1} \langle s, \gamma_{m,l} \rangle e^{-j2\pi\left(\frac{k}{K}m-\frac{n}{L}l\right)} = \\ \sqrt{KL} Z_s[n,k]  Z_{\gamma}^{\ast}[n,k],
\end{multline}
which follows form \eqref{eq:DZTproduct}. Using \eqref{eq:Zs} we get
\begin{multline}
    \frac{1}{\sqrt{KL}} \sum_{m=0}^{K-1}\sum_{l=0}^{L-1} \langle s, \gamma_{m,l} \rangle e^{-j2\pi\left(\frac{k}{K}m-\frac{n}{L}l\right)} = \\ KL Z_x[n,k]Z_g[n,k] Z_{\gamma}^{\ast}[n,k].
    \label{eq:overlay_chain}
\end{multline}
From \eqref{eq:overlay_chain} it is apparent that the overlay is equivalent to the scaled elementwise product of the DZTs $Z_x$, $Z_g$, and $Z_{\gamma}^{\ast}$. To obtain the transmit symbols $Z_x$ we require that the product $Z_g[n,k] Z_{\gamma}^{\ast}[n,k] = 1/(KL)$, which is equivalent to the the biorthogoanlity condition in \eqref{eq:biorthogonality}. Thus, even if nonrectangular pulses are used, the resulting processing chain of pulse shaping OFDM modulation and demodulation is equivalent to simply calculating the IDZT and DZT.

\begin{remark}{}In the presence of a channel, additional effects need to be considered, which is out of the scope of this paper. Analysis for pulses of length $L$ is presented in \cite{Raviteja2019}. As mentioned in \cite{Raviteja2019}, nonrectangular pulses lead to a nonconstant channel gain of the channel in the DD domain.
\end{remark}

%% file: Figures/overlay.tex
\tikzstyle{block} = [rectangle, draw, rounded corners, minimum height=3em,  align = center]
\tikzstyle{block1} = [rectangle, draw, text width=0.8cm,text centered, rounded corners, minimum height=2em,  align = center]
\tikzstyle{block2} = [rectangle, draw, text width=1.1cm,text centered, rounded corners, minimum height=2em, , align = center]
\tikzstyle{line} = [draw, -latex']
\tikzstyle{line1} = [draw, latex'-]
    
\begin{tikzpicture}[node distance = 2.5cm, line width = .75pt]
 \tikzstyle{every node}=[font=\tiny]

\begin{scope}[local bounding box=scope1, draw, align = center, node distance = 0.8cm, auto]
\node[block2](g0){$g_{m,0}[n]$};
\node[block2, below of = g0] (g1) {$g_{m,1}[n]$};
\node[below of = g1] (gx){$\vdots$};
\node[block2, below of = gx] (gL1) {$g_{m,L-1}[n]$};
\node[below of = g1, yshift = 0.75cm, xshift =1.25cm, circle, align = center, draw, inner sep = 0.5pt] (adder) {\scriptsize $+$};

\path[line] (g0) --++ (0.8,0) -- (adder);
\path[line] (g1) --++ (0.8,0) -- (adder);
\path[line] (gL1) --++ (0.8,0) -- (adder);

\begin{scope}[node distance = 1.1cm]
\node[left of = g0, anchor=east ] (X_0) {$a_{m,0}$};
\node[left of = g1, anchor=east] (X_1) {$a_{m,1}$};
\node[left of = gL1, anchor=east] (X_L1) {$a_{m,L-1}$};
\end{scope}

\path[line] (X_0)--(g0);
\path[line] (X_1)--(g1);
\path[line] (X_L1)--(gL1);

\node[draw, rectangle, rounded corners, xshift = 1.1cm, inner sep = 2pt , align = center] (channel) at (adder) {discrete\\TF-dispersive\\channel};
\path[line] (adder)--(channel);
\node[anchor = south] at ($(adder.east)!0.35!(channel.west)$){$s$};
\coordinate[xshift=0.9cm] (sep) at (channel){};
\draw[] (channel)--(sep);

\node[anchor = south] at ($(channel.east)!0.5!(sep.west)$){$r$};

\node[block2, above of = sep, yshift = -0.75cm, xshift =1.25cm] (gamma1) {$\gamma_{m,1}[n]$};
\node[block2, above of = gamma1] (gamma0) {$\gamma_{m,0}[n]$};
\node[below of = gamma1] (gammax){$\vdots$};
\node[block2, below of = gammax] (gammaL1) {$\gamma_{m,L-1}[n]$};

\path[line1] (gamma0) --++ (-1,0) -- (sep);
\path[line1] (gamma1) --++ (-1,0) -- (sep);
\path[line1] (gammaL1) --++ (-1,0) -- (sep);

\begin{scope}[node distance = 1.1cm]
\node[right of = gamma0, anchor = west] (Y_0) {$\hat{a}_{m,0}$};
\node[right of = gamma1, anchor = west] (Y_1) {$\hat{a}_{m,1}$};
\node[right of = gammaL1, anchor = west] (Y_L1) {$\hat{a}_{m,L-1}$};
\end{scope}

\path[line] (gamma0)--(Y_0);
\path[line] (gamma1)--(Y_1);
\path[line] (gammaL1)--(Y_L1);
\end{scope}

\begin{scope}[node distance = 2.25cm,shift={($(scope1.north)+(0,2.25cm)$)}]
\node[block, font = \footnotesize] (transmission) { Pulse Shaping\\OFDM \\Transmission};
\node[block1, left of = transmission] (isfft){\footnotesize ISFFT};
\node[left of = isfft, xshift = 1cm, anchor = east] (source_DD){\footnotesize$\Zxs$};
\path[line] (source_DD)--(isfft);
\path[line] (isfft)--(transmission);
\node[anchor = south]at ($(isfft.east)!0.5!(transmission.west)$) {\footnotesize$a$};
\node[block1, right of = transmission] (sfft){\footnotesize SFFT};
\path[line] (transmission)--(sfft);
\node[anchor = south] at ($(transmission.east)!0.5!(sfft.west)$) {\footnotesize$\hat{a}$};
\node[right of = sfft, xshift = -1cm, anchor = west] (rx) {\footnotesize$\Zys$};
\path[line] (sfft)--(rx);
\end{scope}

\coordinate (box1) at ($(X_0)!0.4!(g0)+(0,0.75cm)$);
\coordinate (box2) at ($(gammaL1)!0.5!(Y_L1)+(0,-0.75cm)$);
\coordinate (box3) at ($(gamma0)!0.5!(Y_0)+(0,0.75cm)$);

\draw[dashed] (transmission.south west)--(box1);
\draw[dashed] (transmission.south east)--(box3);

\draw[rounded corners] (box1) rectangle (box2);
 
\node[] at (-2.,3.25){\footnotesize \textbf{a)}};
\node[] at (-2.,1.5) {\footnotesize \textbf{b)}};
\end{tikzpicture}

%% file: OTFS.tex
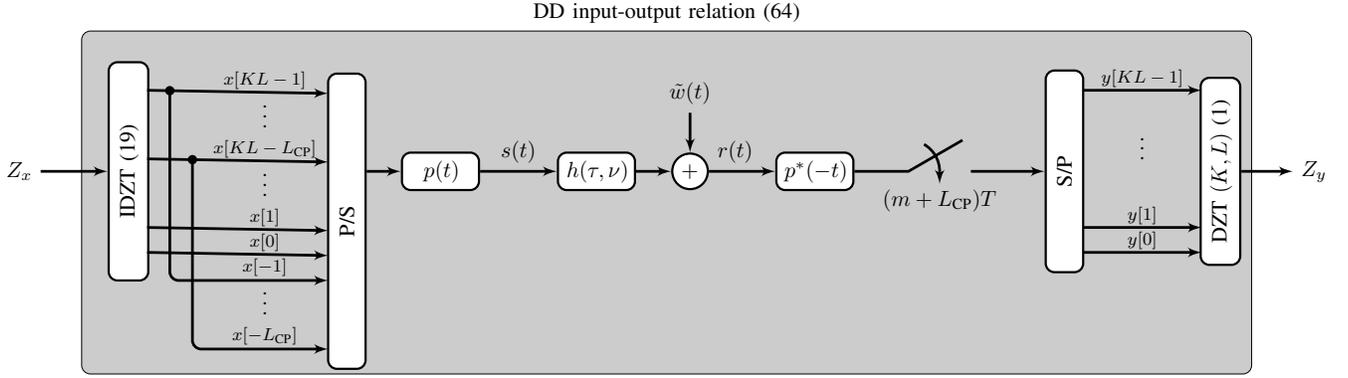
\begin{figure*}[t!]
    \centering
    \resizebox{2.05\columnwidth}{!}{\input{Figures/OTFS_system}}
    \caption{OTFS system model considered in this work. The IDZT maps a sequence the symbols defined in the DD domain to a discrete sequence. A CP is added by copying the last $\Lcp$ samples. The resulting sequence $x$ is converter to a serial stream by a parallel-to-serial converter (P/S) before mapped onto a pulse $p(t)$ and send over a noisy TF-dispersive channel $h(\tau,\nu)$. At the receiver, a sampled matched filter is applied before the serial stream is converter to a parallel stream by a serial-to-parallel (S/P) converter. Lastly, the sequence $y$ is mapped to the DD domain using the DZT.}
    \label{fig:OTFS_system}
\end{figure*}
In Sec.~\ref{sec:overlay}, we described OTFS as an overlay for OFDM and showed that the modulation/demodulation is equivalent to the IDZT/DZT. Here, we will use the IDZT/DZT to map the symbols in the DD domain directly to a time domain sequence and vice versa. We further consider a pulse-amplitude modulation (PAM) system to map the discrete symbols onto continuous pulses, as schematically shown in Fig.~\ref{fig:OTFS_system}. This approach allows the digital implementation of OTFS similar to the PAM implementation of OFDM as presented in \cite[Chapter~6.4.2.]{barry2004digital}, \cite{Gorokhov2004} or \cite{Luise1996}. 

\subsection{Transmitter}
At the transmitter, a frame $\Zxs$ is mapped to a sequence $x$ using the IDZT in \eqref{eq:IDZT2}. To ensure a cyclic behavior of the delayed signal and avoid interference between consecutive frames, a cyclic prefix (CP) of length $\Lcp$ is added. The choice of the length of the CP will be addressed later. The CP is added by copying the last $\Lcp$ samples and inserting them at the beginning of the sequence (see Fig.~\ref{fig:OTFS_system}). The elements of the sequence $x$ are then mapped onto time-shifted pulses $p(t)$ using PAM. The transmitted signal is given as:
\begin{equation}
        s(t) = \sum_{n=0}^{N+\Lcp-1} x[n-\Lcp] p(t-nT),
        \label{eq:transmittSignal}
\end{equation}
where $T$ is the modulation interval and $p(t)$ is a square-root Nyquist pulse. 

\begin{remark}{}
When choosing the frame size $K$ and $L$, we implicitly define the period of the sequence $x$. However, for a fixed period, $K$ and $L$ are not defined unambiguously. Different choices of $K$ and $L$ modify the behavior of the overall system. To see this, consider first the case $K=1$. The symbol of $\Zxs$ are arranged on a line along the delay axis. The IDZT does not alter the sequence and can be skipped, see \eqref{eq:dzt_K1}. The system is a single carrier system. On the other hand, for $L=1$, the symbols $\Zxs$ are arranged along the Doppler axis. The IDZT is simply the IDFT (see \eqref{eq:IDFT}), and  \eqref{eq:transmittSignal} becomes an OFDM system as in \cite{Luise1996} or \cite{Gorokhov2004}. 
\end{remark}

\subsection{Channel Model  Input-Output Relation}
We consider TF dispersive channels and model the received signal as \cite[Sec.~1.3.1]{Hlawatsch}
\begin{equation}
    r(t) = \int_{-\infty}^{\infty}\int_{-\infty}^{\infty} h(\tau,\nu) s(t-\tau)e^{j2\pi\nu t } d\tau d\nu + \tildew(t)
    \label{eq:r_tf_dispersive}
\end{equation}
where $h(\tau,\nu)$ is the so-called DD spreading function. The complex noise $\tildew(t)$ is assumed to be white and Gaussian with power spectral density $N_0$.
We model the channel by $P$ discrete scattering objects. Each scattering object is associated with a path delay $\tau_p$, a Doppler shift $\nu_p$, and a complex attenuation factor $\alpha_p$. Thus, the spreading function $h(\tau,\nu)$ becomes 
\begin{equation}
    h(\tau,\nu) = \sum_{p=0}^{P-1} \alpha_p\delta(\tau-\tau_p)\delta(\nu-\nu_p).
    \label{eq:channel}
\end{equation}
Substituting \eqref{eq:channel} in \eqref{eq:r_tf_dispersive} yields:
\begin{equation}
    r(t) = \sum_{p=0}^{P-1} \alpha_p s(t-\tau_p)e^{j2\pi\nu_p t} +\tildew(t),
    \label{eq:rx}
\end{equation}
i.e., a superposition of scaled, delayed, and Doppler shifted replicas of the transmitted signal. The Doppler shift is given by $\nu_p = v_p \fc/c$, where $v_p$, $\fc$ and $c$ are, respectively, the relative velocity of the $p$th scattering object, the  carrier frequency, and the speed of light. The CP's length in \eqref{eq:transmittSignal} is chosen such that $\Lcp T$ is larger than or equal to the maximum delay.

Using \eqref{eq:transmittSignal} in \eqref{eq:rx} gives
\begin{multline}
    r(t) = \sum_{p=0}^{P-1} \alpha_p \sum_{n=0}^{N+\Lcp-1} x[n-\Lcp] p(t-nT-\tau_p)e^{j2\pi \nu_p t}\\ + \tildew(t).
\end{multline}

\subsection{Receiver}
At the receiver, a matched filter with impulse response $p^{\ast}(-t)$ is applied. The output of the matched filter $y(t)$ is 
\begin{multline}
    y(t) = \sum_{p=0}^{P-1} \alpha_p \sum_{n=0}^{N+\Lcp-1} x[n-\Lcp] \\ 
     \int_{-\infty}^{\infty}   p(\tau-nT-\tau_p)e^{j2\pi \nu_p \tau }  p^{\ast}(\tau-t) d\tau + w(t),
         \label{eq:mf}
\end{multline}
where $w(t)$ is the filtered noise. Assuming that the pulse bandwidth is much larger than the maximum Doppler shift, we can approximate the integral in \eqref{eq:mf} as
\begin{multline}
  \int_{-\infty}^{\infty}  p(\tau-nT-\tau_p)e^{j2\pi \nu_p \tau }  p^{\ast}(\tau-t) d\tau  \approx \\
  e^{j2\pi\nu_p (nT+\tau_p)} h(t-nT-\tau_p),
  \label{eq:ambiguity_function}
\end{multline}
where 
\begin{equation}
h(t) = \int_{-\infty}^{\infty}  p(\tau) p^{\ast}(\tau-t)d\tau 
\end{equation}
is the corresponding Nyquist.
Using approximation \eqref{eq:ambiguity_function}, the output of the matched filter is 
\begin{multline}
y(t) \approx \sum_{p=0}^{P-1} \alpha_p \sum_{n=0}^{N+\Lcp} x[n-\Lcp] e^{j2\pi \nu_p (n T+\tau_p)}  \\  h(t-nT-\tau_p) + w(t).
\end{multline}

The matched filter output is sampled every $T$ seconds and with an offset of $\Lcp T$ to discard the CP. The sampled signal ${y[m]= y((m+\Lcp)T)}$ is
\begin{align}
          y[m]  = \sum_{p=0}^{P-1} \alpha_p \sum_{n=-\Lcp}^{N-1}   x[n] e^{j2\pi \frac{k_p}{KL} n} h_p[m-n] +w[m],
            \label{eq:linConv}
\end{align}
where $h_p[n] = h(nT-\tau_p)$ is the sampled pulse and $w[m]$ are independent and identically distributed (i.i.d.) complex Gaussian random variables with variance $N_0$. To shorten the notation, we combined the constant phase terms with the channel gain. Furthermore, we express $\nu_p$ as a multiple of the Doppler resolution which we define as
\begin{equation}
\delnu \triangleq \frac{1}{KLT},
\label{eq:Doppler_res}
\end{equation}
i.e., $\nu_p = \delnu k_p$. The choice of this substitution will be addressed in Sec.~\ref{subsec:Doppler}.

We can bound the interval for which $h(t)$ is significantly different from zero (for sufficient large $L$) to $\pm LT/2$. Thus, we can express $h_p[n]$ as
\begin{equation}
    h_p[n]=
    \begin{cases} h(nT-\tau_p), &\text{for}  -\frac{LT}{2}\leq  nT-\tau_p < \frac{LT}{2}, \\
    0, &\text{else}.
    \end {cases}
    \label{eq:pulse_per}
\end{equation}

\begin{myExample}[Nyquist Pulse]\label{ex:pulse_periodic}
A typical choice of a square-root Nyquist pulse $p(t)$ is the square-root raised cosine pulse \cite[eq.~5.60]{barry2004digital}. The corresponding Nyquist pulse $h(t)$ is the raised cosine pulse \cite[eq.~5.8]{barry2004digital} given as
\begin{equation}
    h(t) = \frac{\sin(\pi t/T)}{\pi t/T} \frac{\cos(\beta \pi t/T)}{1-(2\beta t/T)^2}.
\end{equation}
The sequence $h_p$ that is obtained from sampling a raised cosine pulse with roll-off factor $\beta=0.5$ and that is delayed by $\tau_p=0.5T$ is presented in Fig.~\ref{fig:hp}. It can be seen that the samples are only significantly different from zero for $n\in [-3,4]$. 
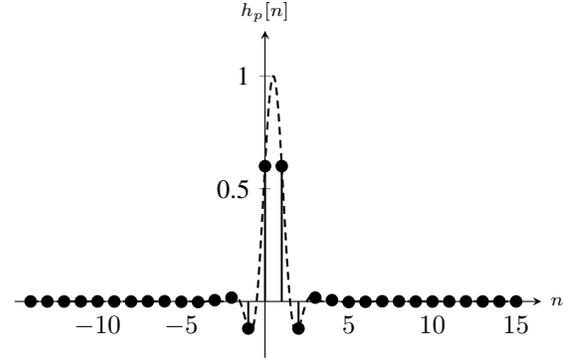
\begin{figure}
    \centering
    \input{Figures/hp}
    \caption{Example of a sequence $h_p$ obtained by sampling a fractionally delayed $(\tau=0.5T)$ raised cosine pulse with roll-off factor $\beta=0.5$, indicated by the dashed line.}
    \label{fig:hp}
\end{figure}
\end{myExample}

The CP allows to approximate the linear convolution in \eqref{eq:linConv} by a circular convolution. The sample $y[m]$ is then 
\begin{equation}
 y[m]  =\sum_{p=0}^{P-1} \alpha_p y_p[m] + w[m],
 \label{eq:circConv}
\end{equation}
where
\begin{equation}
    y_p[m] \approx \sum_{n=0}^{KL-1}   x[n] e^{j2\pi \frac{k_p}{N} n} h_p[m-n].
    \label{eq:y_p}
\end{equation}
Here, $h_p$ is periodized with period $KL$, i.e., ${h_p[n]=h_p[n+KL]}$. Note that if all delays $\tau_p$ coincide with a integer multiple of the sampling period, i.e., $\tau_p = n_pT$ with $n_p\in\mathbb{N}$, then $h_p[n]=\delta[n-n_p]$  and \eqref{eq:y_p} is exact.

The receiver calculates the DZT of the sequence $y$ (see Fig.~\ref{fig:OTFS_system}). The DZT is a linear transform, and thus, we can study the effect of the individual delay and Doppler shifts. Therefore, we write the DZT of $y$ in \eqref{eq:circConv} as the superposition of the delayed and Doppler-shifted signals
\begin{equation}
  \Zys[n,k] =  \sum_{p=0}^{P-1} \alpha_p  \Zyp[n,k] + Z_w[n,k],
  \label{eq:DD_rx}
\end{equation}
where $\Zyp$ is the DZT of the sequence $y_p$ in \eqref{eq:y_p} and $Z_w$ is the DZT of the sampled noise sequence $w$.

The sequence $y_p$ in \eqref{eq:y_p} can be recognized as the circular convolution of the modulated signal $x[n]e^{j2\pi k_pn/N}$ with the sequence $h_p$, i.e.,
\begin{equation}
  y_p  =\left(x \cdot u_p\right)\circledast h_p,
\end{equation}
where $u_p[n] =e^{j2\pi (k_p/N) n} $. Using the modulation property \eqref{eq:modProp} and the convolution property \eqref{eq:convProp}, $\Zyp$ is
\begin{multline}
  \Zyps[n,k] = \sum_{m=0}^{L-1}
  \left(\sum_{l=0}^{K-1} Z_x[m,l] Z_{u_p}[m,k-l]\right) \\
  Z_{h_p}[n-m,k],
  \label{eq:dzt_conv}
\end{multline}
i.e., the convolution of $Z_x$ with the DZTs $\Zups$ and $\Zh$ of the sequences $u_p$ and $h_p$, respectively.

From \eqref{eq:dzt_conv}, it is apparent that the inner convolution causes a spread of symbols $Z_x$ with respect to $k$ (Doppler spread), whereas the outer convolution with respect to $n$ (delay spread). The spread of the symbols of $\Zxs$ causes interference with adjacent symbols in the delay Doppler domain. We evaluate the spread in the individual domains in Sec.~\ref{sec:DD}.

To complete the evaluation of \eqref{eq:DD_rx}, we need to evaluate the noise DZT $Z_w$, which can be done as follows. Since we assume a square-root Nyquist filter $p^{\ast}(-t)$, it follows that the noise samples $w[n]$ are i.i.d. and distributed according to ${\mathcal{CN}(0,N_0)}$. Since the DZT $Z_w[n,k]$ for a given $n$ is the unitary DFT of the subsampled noise sequence $w^{(n,L)}$, we conclude that the elements of $Z_w$ share the same statistics as the elements of the sequence $w$, i.e., ${Z_w[n,k]\sim\mathcal{CN}(0,N_0)}$. 
 
We conclude this section by noting that the spread of the symbols causes intersymbol interference (ISI) in the DD domain and a symbol detector has to cope with this ISI to obtain an estimate of the transmitted symbols $\Zxs$. Treatment of the symbol detection is out of the scope of this paper. Examples of symbol detectors based on iterative message passing algorithms are presented in \cite{Raviteja} and \cite{Gaudio}. 

%% file: Figures/OTFS_system.tex
\tikzstyle{block} = [rectangle, draw, text width=1cm, text centered, rounded corners, minimum height=.6cm, fill=white]
\tikzstyle{dot} = [circle, fill, minimum size=0.15cm, inner sep=0pt, outer sep=0pt]
\tikzstyle{line} = [draw, -latex', line width = 1.25pt, rounded corners]

\begin{tikzpicture}[node distance = 2.5cm, auto, line width = 1pt]


\node[] (Zx) {$Z_x$};

\node[rectangle, xshift = -0.75cm, draw, text width=.4cm, text centered, rounded corners, minimum height=3.5cm,, right of = Zx,  fill = white ] (idzt){\rotatebox{90}{IDZT \eqref{eq:IDZT2}}};
\path[line] (Zx)--(idzt);



\node[rectangle, draw, text centered, rounded corners, right of = idzt, text width=4.5cm,xshift = 1cm,yshift = -0.8cm, rotate = 90, minimum height=0.6cm, fill = white] (cp){P/S};

\coordinate (N1) at ($(idzt)+(0.3,1.3)$);
\coordinate (N2) at ($(cp)+(-0.3,2.05)$);
\path[line] (N1)--(N2);
\node[yshift = 1.9cm] (xN1) at (cp) {};

\coordinate (N3) at ($(idzt)+(0.3,0.2)$);
\coordinate (N4) at ($(cp)+(-0.3,0.95)$);
\path[line] (N3)--(N4);
\node[yshift = 1.1cm] (xNCP) at (cp) {};

\coordinate (N5) at ($(idzt)+(0.3,-1.3)$);
\coordinate (N6) at ($(cp)+(-0.3,-0.55)$);
\path[line] (N5)--(N6);
\node[yshift = -0.7cm] (x1) at (cp) {};

\coordinate (N7) at ($(idzt)+(0.3,-0.9)$);
\coordinate (N8) at ($(cp)+(-0.3,-0.15)$);
\path[line] (N7)--(N8);
\node[yshift = -0.3cm] (x2) at (cp) {};

\node[rotate = 90]  at ($(x2)!0.5!(xNCP)$) {};

\coordinate (N9) at ($(N1)!0.125!(N2)$);
\coordinate (N10) at ($(N6)+(0,-0.4)$);
\path[line] (N9)|-(N10);
\node[dot] at (N9){};
\node[yshift = -1.1cm] (x11) at (cp) {};

\coordinate (N11) at ($(N3)!0.25!(N4)$);
\coordinate (N12) at ($(N10)+(0,-1.1)$);
\path[line] (N11)|-(N12);
\node[dot] at (N11){};
\node[yshift = -1.9cm] (x12) at (cp) {};

\coordinate (N14) at ($(N5)-(0,0.4)$);
\coordinate (N15) at ($(N14)-(0,1.1)$);

\node[anchor = south, yshift = -0.08cm] (KL) at ($(N1)!0.65!(N2)$){\footnotesize$x[KL-1]$};
\node[anchor = south, yshift = -0.08cm] (KLminus) at ($(N3)!0.65!(N4)$){\footnotesize$x[KL-\Lcp]$};
\node[rotate = 90] at ($(KL)!0.5!(KLminus)$){$\dots$}; 

\node[anchor = south, yshift = -0.08cm] at ($(N5)!0.65!(N6)$){\footnotesize$x[0]$};
\node[anchor = south,yshift = -0.08cm] (one) at ($(N7)!0.65!(N8)$){\footnotesize$x[1]$};
\node[rotate = 90] at ($(one)!0.5!(KLminus)$){$\dots$}; 
\node[anchor = south,yshift = -0.08cm] (minus1) at ($(N14)!0.65!(N10)$){\footnotesize$x[-1]$};
\node[anchor = south,yshift = -0.08cm] (minusL) at ($(N15)!0.65!(N12)$){\footnotesize$x[-\Lcp]$};

\node[rotate = 90] at ($(minus1)!0.5!(minusL)$){$\dots$};

\node[block, xshift = -1cm, yshift = 0.8cm, right of = cp] (pulseshaping) {$p(t)$};
\coordinate[yshift = 0.8cm] (cp1) at (cp.south);
\path[line] (cp1)--(pulseshaping);

\node[block,  right of = pulseshaping] (channel) {$h(\tau,\nu)$};
\path[line] (pulseshaping)--(channel);
\node[anchor = south]at($(pulseshaping)!0.5!(channel)$){$s(t)$};
 \node[circle, draw ,right of = channel, xshift = -1cm, anchor=center, inner sep = 0.075cm, fill = white] (adder) {$+$};
\path[line] (channel)--(adder); 
\node[above of =  adder, yshift = -1.25cm] (noise) {$\tildew(t)$};
\path[line] (noise)--(adder);

\node[block, xshift = -0.5cm,right of = adder] (mf) {$p^{\ast}(-t)$};
\path[line] (adder)--(mf); 
\node[anchor=south]at($(adder.west)!0.35!(mf.east)$){$r(t)$}; 

\coordinate (sampler1) at ($(mf)+(1.5,0)$);

\path[draw, line width = 1.25pt] (mf)--(sampler1)--++(30:1); 

\coordinate (sampler3) at ($(sampler1)+(0.5,0)$);
\draw[-latex', line width = 1.25pt] (sampler3)--++(0,-0.2);

\node[anchor = north, yshift =-0.15cm] at (sampler3){$(m+\Lcp)T$};

\draw[line width = 1.25pt] (sampler3) arc (0:60:0.5);

\coordinate (sampler2) at ($(sampler1)+(1,0)$);
\node[rectangle, xshift = -1cm,  draw, text centered, rounded corners,  right  of = sampler2, text width=3cm, rotate = 90, minimum height=0.6cm, fill = white ] (rmvcp){S/P};
\path[line] (sampler2)--(rmvcp);

\node[yshift=1.3cm] (y0t) at (rmvcp){};
\node[yshift=0.9cm] at (rmvcp){};
\node[yshift=-1.3cm] (y0N1t) at (rmvcp){};

\coordinate (y0) at ($(rmvcp)+(0.3,1.3)$);
\coordinate (y1) at ($(rmvcp)+(0.3,-0.9)$);
\coordinate (yN1) at ($(rmvcp)+(0.3,-1.3)$);

\node[rectangle, draw, text width=.4cm, text centered, rounded corners, minimum height=3cm, right of = rmvcp, fill=white] (dzt){\rotatebox{90}{DZT $(K,L)$ \eqref{eq:DZT}}};

\coordinate (y02) at ($(dzt)+(-0.3,1.3)$);
\coordinate (y12) at ($(dzt)+(-0.3,-0.9)$);
\coordinate (yN12) at ($(dzt)+(-0.3,-1.3)$);

\path[line] (y0)--(y02);
\node[anchor = south, yshift = -0.08cm] (z) at ($(y0)!0.5!(y02)$){\footnotesize$y[KL-1]$};
\path[line] (y1)--(y12);
\node[anchor = south, yshift = -0.08cm] (z1) at ($(y1)!0.5!(y12)$){\footnotesize$y[1]$};
\path[line] (yN1)--(yN12);
\node[anchor = south, yshift = -0.08cm] at ($(yN1)!0.5!(yN12)$){\footnotesize$y[0]$};
\node[rotate = 90] at ($(z)!0.5!(z1)$) {$\dots$};

\node[right of = dzt, xshift = -1cm] (Zy) {$Z_y$};
\path[line](dzt)--(Zy);


\begin{scope}[local bounding box=scope1,on background layer]
\draw[rounded corners, fill = gray!40] ($(Zx)+(1,2.25)$) rectangle  ($(Zy)+(-1,-3.25)$);
\end{scope}
\node[anchor = south] at (scope1.north) {DD input-output relation \eqref{eq:DD_rx}};

\end{tikzpicture} 

%% file: Figures/hp.tex
\begin{tikzpicture}[
declare function={func(\x) = (\x==0)*5.65;}
]
\begin{axis}[
axis x line = center,
axis y line = center,
scale only axis = true,
width = 7cm,
xmin = -14.95, xmax =16.5,
ymin = -.25, ymax = 1.2,
scale only axis = true,
xlabel={$n$},
ylabel={$h_p[n]$},
ytick = {-14:15},
ytick = {0,0.5,1},
yticklabels={{0,0.5,1}},
trig format plots=rad,
grid = none,
y = 3cm,
x label style={anchor=west, font = \scriptsize},
y label style={anchor=south, font = \scriptsize},
]

\addplot[ycomb, ,mark=*, color = black,domain = -14:15, samples = 30, line width = 0.75pt] {
sin(pi*(x-0.5))/(pi*(x-0.5))*cos(pi/2*(x-0.5))/(1-(x-0.5)^2)
};

\addplot[densely dashed, opacity =1, domain = -14:15, samples = 300, smooth, line width = 0.8pt] {
sin(pi*(x-0.5))/(pi*(x-0.5))*cos(pi/2*(x-0.5))/(1-(x-0.5)^2)
};

\end{axis}
\end{tikzpicture}

%% file: DD.tex
In this section, we will evaluate the spread of the symbols in the DD domain given by \eqref{eq:dzt_conv}. We start with the evaluation of the inner convolution which causes a spread of the symbols in the Doppler domain and then proceed with the evaluation of the outer convolution resulting in a spread of a symbol in the delay domain. In the last step, we present the overall spread in the DD domain considering three different channels.

\begin{figure}
    \centering
    \input{Figures/Doppler_interference}
    \caption{$|V[k-k_p]|$ in \eqref{eq:dirichlet} for $K=30$. The magnitude is symmetric, and thus, only the part for which $|k-k_p|\geq 0$ is shown.}
    \label{fig:DopplerSpreading}
\end{figure}
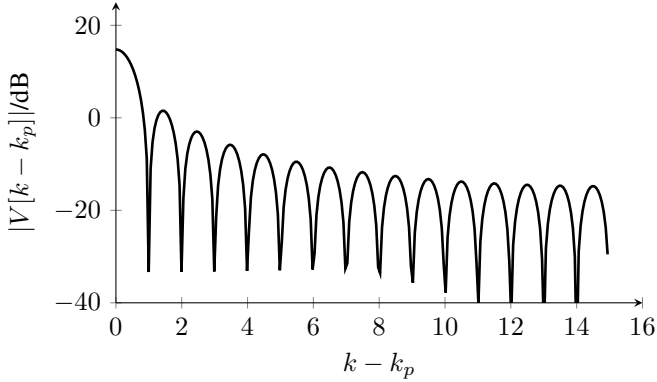

\subsection{Doppler Spread}
\label{subsec:Doppler}
Before evaluating the Doppler spread, we recall that the DZT is periodic in $k$, or equivalently, in the Doppler domain. Since Doppler shifts are, unlike delays, positive or negative, we consider the fundamental rectangle symmetric around $k=0$. The Doppler spread is determined by the DZT $\Zups$ which is 
\begin{align}
    \Zups[n,k]  &= \frac{1}{\sqrt{K}}\sum_{l=0}^{K-1} e^{2\pi\frac{k_p}{KL}(n+lL)}e^{-2\pi\frac{k}{K}l}\\
                &= e^{j2\pi \frac{k_p}{KL}n} V[k-k_p],\label{eq:doppler_spread}
\end{align}
where 
\begin{equation}
V[k-k_p] = \frac{1}{\sqrt{K}}e^{-j\pi\frac{K-1}{K}(k-k_p)} \frac{\sin\left(\pi(k-k_p)\right)}{\sin\left(\frac{\pi}{K}(k-k_p)\right)},
      \label{eq:dirichlet}
\end{equation}
which is also known as Dirichlet or periodic sinc function. The definition of the Doppler resolution \eqref{eq:Doppler_res} becomes clear at this point:
if $k_p$ is an integer, $V[k-k_p]$ reduces to $\sqrt{K} \delta[k-k_p]$ and \eqref{eq:dzt_conv} becomes a translation (up to a complex factor) of the DZT $\Zxs$ by $k_p$ points in the Doppler domain. 

For noninteger values of $k_p$, interference in the Doppler domain as illustrated in Fig.~\ref{fig:DopplerSpreading} occurs. The interference can also be understood as spectral leakage occurring in the DFT analysis of sinusoidal signals whose frequency do not match the discrete frequencies of the DFT \cite{Harris1978}.

\subsection{Delay Spread}
The DZT of the pulse $h_p$ determines the delay spread. We assumed that the support of $h_p$ within one period to be smaller than or equal to $L$ consecutive samples, see \eqref{eq:pulse_per}. Due to this assumption, we can infer that the magnitude $|Z_{h_p}|$ is independent of $k$, as shown in Example~\ref{ex:DZT}. Thus, all symbols experience the same spread in the delay domain. The interference itself is governed by the pulse $h(t)$. Thus, a more concentrated pulse results in less interference in the delay domain. For the case of the raised cosine pulse in Example~\ref{ex:pulse_periodic}, it means that a larger roll-off factor $\beta$ reduces the delay spread. However, a larger roll-off factor comes with the cost of increased bandwidth requirements.

Similar to the Doppler resolution $\delnu$, we can define the delay resolution $\delres$. For a delay $\tau_p=n_pT$ with $n_p \in \mathbb{N}$, $|Z_{h_p}[n,k]|$ reduces to $\delta[n-n_p]$. The outer convolution, therefore, causes a shift of $n_p$ grid points in the delay domain. Thus, we define the delay resolution as the value for which the DZT gets shifted by one sample in the delay, which is
 \begin{equation}
     \delres \triangleq T.
 \end{equation}

\begin{myExample}[Delay Spread]\label{ex:Delay_spread}
Consider the sequence $h_p$ shown in Fig.~\ref{fig:hp}. The DZT of the sequence $Z_{h_p}$ is illustrated in Fig.~\ref{fig:DelaySpreading}. It can be seen that the magnitude $|Z_{h_p}[n,k]|$ is independent of $k$ and the phase $\varphi_{h_p}[n,k]$ shows an additional linear phase for $n \geq 16$.

\begin{figure}
\includegraphics[]{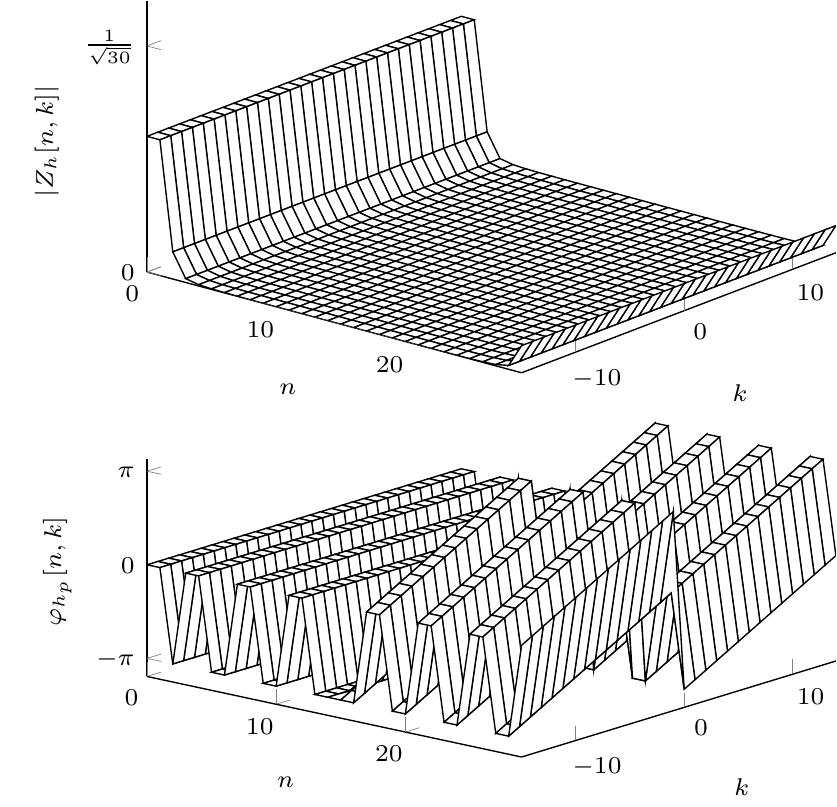}
\caption{The DZT $Z_{h_p}$ of the sequence $h_p$ in Fig.~\ref{fig:hp}. The magnitude response is constant for $k$ due to the limited support of the pulse within one period.}
\label{fig:DelaySpreading}
\end{figure}
\end{myExample}

\subsection{Delay-Doppler Spread}
Having established the individual spreading functions $\Zups$ and $\Zh$, we provide two introductory examples for the overall spread in the DD domain and one example based on a tapped delay line (TDL) model.
\begin{myExample}[Integer Shifts]
\label{ex:integer_shift}
Consider the particular case of integer shifts in the DD domain, i.e, $k_p\in\mathbb{Z}$ and $\tau_p=n_pT$ with $n_p\in \mathbb{N}$. For this case we have that $\Zups[n,k]=\sqrt{K} e^{j2\pi\frac{k_p}{N}n} \delta[k-k_p]$ and $\Zh[n,k]=(1/\sqrt{K})\delta[n-n_p]$. Thus, \eqref{eq:DD_rx} becomes
\begin{equation}
    \Zys[n,k]= \sum_{p=0}^{P-1} \alpha_p  e^{j2\pi\frac{k_p}{N}n_p}\Zxs[n-n_p,k-k_p],
\end{equation}
i.e., a superposition of scaled and DD-shifted replicas of the transmitted symbols. Recalling from the periodicity properties of the DZT, we see that the interference in the DD domain is cyclic. However, in the delay domain, the \emph{quasi}-periodicity has to be considered. This result coincides with \cite[Eq.~(30)]{Raviteja2019}, where OTFS as an overlay for an OFDM system with rectangular pulses is considered. 
\end{myExample}

\begin{remark}{}
In the OTFS literature, the delay domain is usually subdivided into two parts: indices $n$ for which $nT$ is smaller than the delay $
\tau_p$, and indices greater than or equal to the delay $\tau_p$, see for instance \cite[Sec.~III]{Gaudio} or \cite[Sec.~III]{Raviteja2019}. For indices $n$ with the property $nT < \tau_p$, an additional complex factor is introduced. However, considering OTFS by means of the DZT, we immediately recognize this complex factor given due to the \emph{quasi}-periodicity of the DZT. 
\end{remark}

\begin{myExample}[Fractional Shifts]\label{ex:frac_delay_Doppler}
\begin{figure}
    \centering
    \includegraphics[]{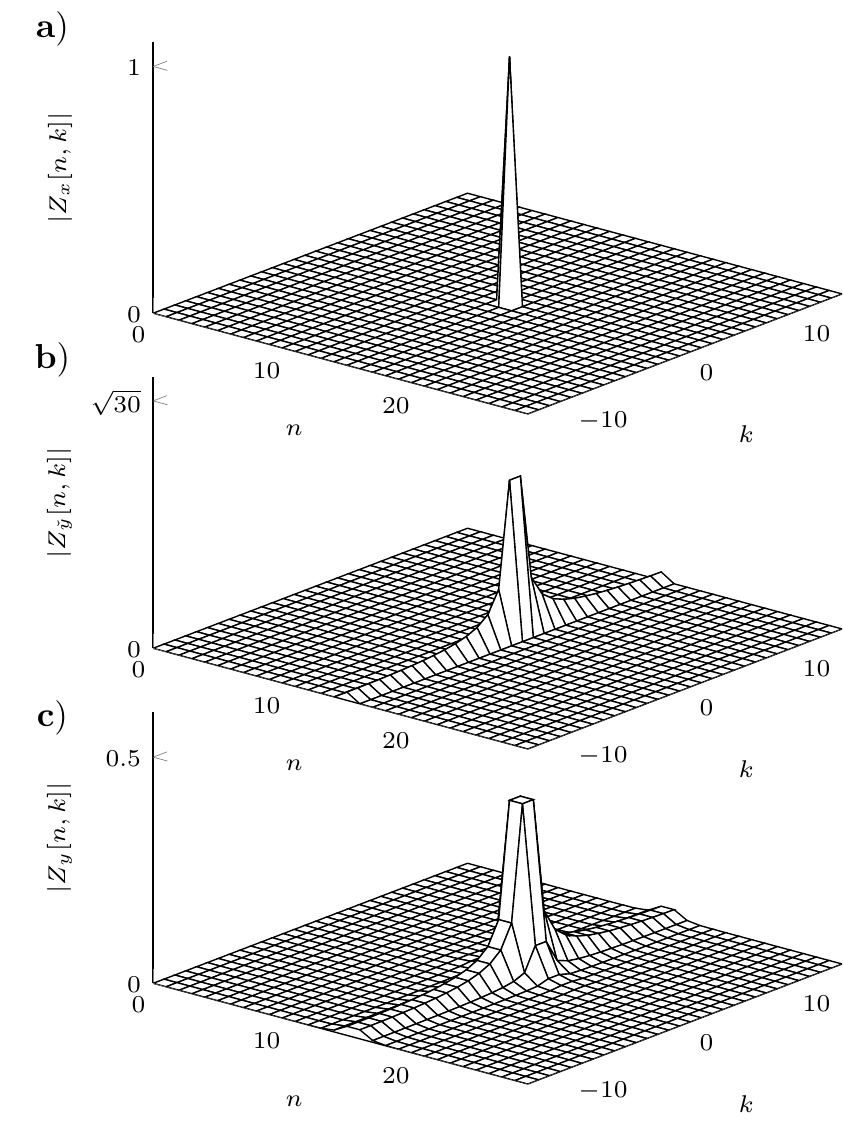}
    \caption{Example of a spread of a symbol a) in the DD domain due to fractional delay and Doppler shift in Example~\ref{ex:frac_delay_Doppler}. The spread can be first evaluated in the Doppler domain b) before it is further spread in the delay domain c).}
    \label{fig:dd_spread}
\end{figure}
In this example, we do not restricted to $k_p$ and $n_p$ to integers. Instead, we present an example causing a maximum spread of a symbol in the DD domain. Therefore, we consider $k_p=0.5$ and $\tau_p = 0.5T$ as shown in Fig.~\ref{fig:DopplerSpreading} and \ref{fig:DelaySpreading}, respectively. For illustrative purposes, we assume only one nonzero element of $\Zx$ at the transmitter, e.g., $\Zx = \delta[n-L/2]\delta[k]$ for $K=L=30$, which is illustrated in Fig.~\ref{fig:dd_spread}a). 

The spread in the DD domain can be evaluated first for the Doppler and then for the delay domain. Therefore, let $Z_{\check{y}}$ be the result of the inner convolution of \eqref{eq:dzt_conv}, i.e., 
\begin{equation}
    Z_{\check{y}}[n,k] = \sum_{l=0}^{K-1} Z_x[n,l] Z_{u_p}[n,k-l].
\end{equation}
The symbol in Fig.~\ref{fig:dd_spread}a) will be spread according to the Doppler spread illustrated in Fig.~\ref{fig:DopplerSpreading} for $k_p=0.5$. The DZT $Z_{\check{y}}$ is shown in Fig.~\ref{fig:dd_spread}b). The outer convolution, which is given by
\begin{equation}
 \Zyps[n,k] = \sum_{m=0}^{L-1} Z_{\check{y}}[m,k] Z_{h_p}[n-m,k],
\end{equation}
causes a spread in the delay domain according to $Z_{h_p}$. In Fig.~\ref{fig:dd_spread}c), the resulting spread in the DD domain is illustrated for $Z_{h_p}$ presented in Example~\ref{ex:Delay_spread}. The symbol initially located on a single point in the DD domain is spread in the DD domain. In particular, the symbol is spread over the entire Doppler domain, whereas the spread in the delay domain is restricted to the vicinity of the original location of the symbol.
\end{myExample}

\begin{figure*}
\begin{minipage}[t]{0.5\textwidth}
\includegraphics[]{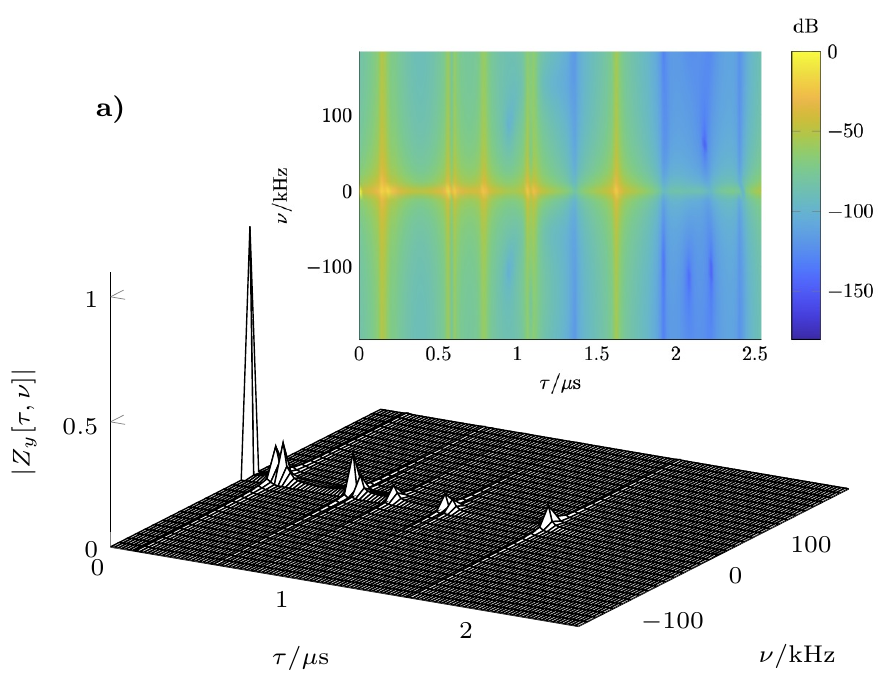}
\end{minipage}
\begin{minipage}[t]{0.5\textwidth}
\includegraphics[]{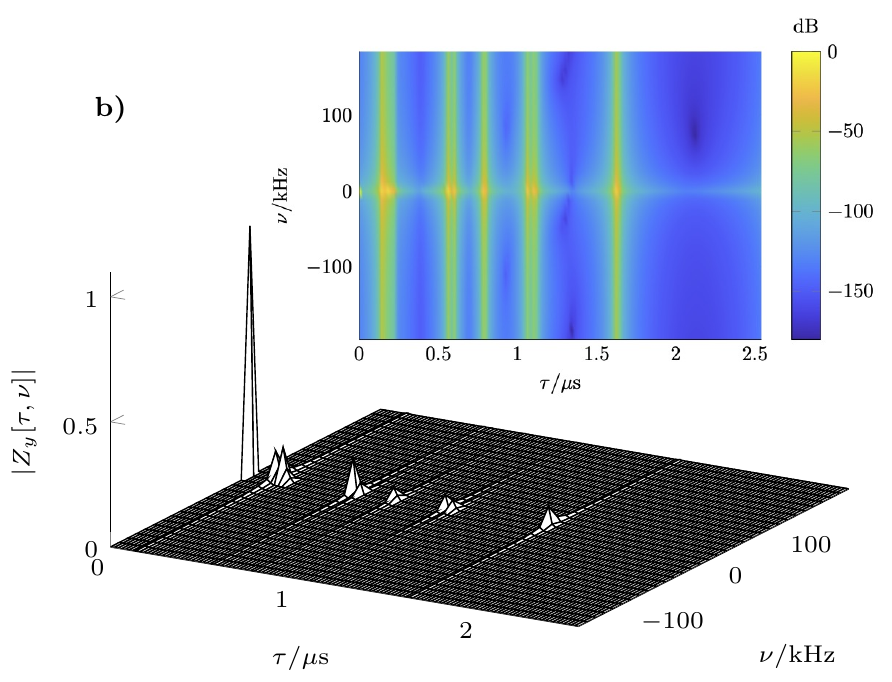}
\end{minipage}
\caption{Example of the DD spread of a TDL-E channel model for a raised cosine pulse $h(t)$ with roll-off factor a) $\beta=0.1$ and b) $\beta = 1$. The channel is composed by a LOS component and 13 additional scattering objects. For illustration purposes, we show $|\Zys|$ in dB in the insets. In both cases, the spread in the Doppler dimension is equivalent. However, the spread in the delay domain is  significantly lower for $\beta = 1$.}
\label{fig:tdlE}
\end{figure*}

\begin{myExample}[TDL-E]
In this concluding example, we provide a DD spread based on a TDL model. The TDL model is suggested for link-level evaluation in 5G \cite[Sec. 7.7.2]{TDLChannel}. The TDL-E model, in particular, assumes a line-of-sight (LOS) path and 13 additional scattering paths. We consider a root mean square delay spread of 300~ns, which is representative of rural areas. Unlike the TDL model suggests, we model each tap by a single Doppler frequency. The Doppler frequency is chosen randomly from a uniform distribution by considering a maximum relative velocity of $\pm150$~km/h and a carrier frequency of 28~GHz. Thus, the maximum Doppler shift is $\nuMax=3888$~Hz. We assume a sampling rate of 50~MHz, a frame of size $L=128$ and $K=32$, and a square-root raised cosine pulse $p(t)$. To visualize the DD spread of the channel, we define only one nonzero symbol in the DZT domain at $n=0$ and $k=0$. The resulting DD spread is visualized for a roll-off factor $\beta = 0.1$ in Fig.~\ref{fig:tdlE}a) and for $\beta = 1$ in Fig.~\ref{fig:tdlE}b). We assume that the receiver is synchronized in both time and frequency to the LOS component. Thus, the LOS component is visible $\tau = 0$ and $\nu=0$. From the figures it can be seen that the interference in the delay domain decreases as the roll-off factor is increased. Furthermore, it is apparent from both figures that the ISI is dominant in the delay domain because $\nuMax<\delnu$.
\end{myExample}

%% file: Figures/Doppler_interference.tex
\begin{tikzpicture}

\begin{axis}[%
width = 7cm,
height = 4cm,
scale only axis,
axis x line = bottom,
axis y line = left,
unbounded coords=jump,
xlabel = {$k-k_p$},
ylabel = {$|V[k-k_p]|$/dB},
xmin=0,
xmax=16,
xtick = {0,2,...,15},
ymin=-40,
ymax=25,
legend style={legend cell align=left, align=left, draw=white!15!black}
]
\addplot [line width = 1pt]
  table[row sep=crcr]{%
0	14.7712125471966\\
0.05859375	14.7221585471516\\
0.1171875	14.5743244625914\\
0.17578125	14.3256495571717\\
0.234375	13.9725443531812\\
0.29296875	13.5096332455018\\
0.3515625	12.9293370194197\\
0.41015625	12.2212182865316\\
0.46875	11.3709520142895\\
0.52734375	10.3586668071729\\
0.5859375	9.15616297212971\\
0.64453125	7.7219794601035\\
0.703125	5.99196963487575\\
0.76171875	3.85938681527994\\
0.8203125	1.12635078107658\\
0.87890625	-2.64346805670107\\
0.9375	-8.79253859029573\\
0.99609375	-33.3440519022942\\
1.0546875	-10.9586084810839\\
1.11328125	-5.24216955611419\\
1.171875	-2.30659997485494\\
1.23046875	-0.526534038944006\\
1.2890625	0.583473325389998\\
1.34765625	1.23021749530764\\
1.40625	1.51607823190757\\
1.46484375	1.4933950970366\\
1.5234375	1.18503106866886\\
1.58203125	0.592001397751391\\
1.640625	-0.305932957025273\\
1.69921875	-1.55628330163543\\
1.7578125	-3.25093122007922\\
1.81640625	-5.57049568565965\\
1.875	-8.91913384702937\\
1.93359375	-14.5155750558084\\
1.9921875	-33.2973656453657\\
2.05078125	-17.32311251997\\
2.109375	-11.0343506445898\\
2.16796875	-7.77744998475387\\
2.2265625	-5.74507441996837\\
2.28515625	-4.41694679404039\\
2.34375	-3.57403324930085\\
2.40234375	-3.108026890277\\
2.4609375	-2.96300780222143\\
2.51953125	-3.11352964809854\\
2.578125	-3.55637379835193\\
2.63671875	-4.3095273345264\\
2.6953125	-5.41767427578871\\
2.75390625	-6.96824535343796\\
2.8125	-9.13291541798391\\
2.87109375	-12.2917243836533\\
2.9296875	-17.5586257991301\\
2.98828125	-33.2193245154131\\
3.046875	-21.3705908882241\\
3.10546875	-14.6147289324027\\
3.1640625	-11.1620241147423\\
3.22265625	-8.99511658189221\\
3.28125	-7.55726382723168\\
3.33984375	-6.61772183339053\\
3.3984375	-6.06308400527583\\
3.45703125	-5.83466951882366\\
3.515625	-5.90519853111612\\
3.57421875	-6.26992433635322\\
3.6328125	-6.94519671538965\\
3.69140625	-7.97345634737014\\
3.75	-9.43830571549809\\
3.80859375	-11.5033559588232\\
3.8671875	-14.5266872039376\\
3.92578125	-19.5298238740501\\
3.984375	-33.10957736298\\
4.04296875	-24.4650777929304\\
4.1015625	-17.232258772644\\
4.16015625	-13.61321533855\\
4.21875	-11.3438827017299\\
4.27734375	-9.82772461025209\\
4.3359375	-8.82165316158136\\
4.39453125	-8.20704455405578\\
4.453125	-7.9224815830701\\
4.51171875	-7.93892368864334\\
4.5703125	-8.25019030626613\\
4.62890625	-8.87110654322735\\
4.6875	-9.84201635045044\\
4.74609375	-11.2429290173348\\
4.8046875	-13.2299075721183\\
4.86328125	-16.1407676124827\\
4.921875	-20.9149400681137\\
4.98046875	-32.9676221396899\\
5.0390625	-27.0554783093248\\
5.09765625	-19.3023591907205\\
5.15625	-15.5244795512749\\
5.21484375	-13.1648680739576\\
5.2734375	-11.5832938609071\\
5.33203125	-10.5235090868944\\
5.390625	-9.86143654097405\\
5.44921875	-9.53285022472173\\
5.5078125	-9.50693694508951\\
5.56640625	-9.77610058132935\\
5.625	-10.3536849999795\\
5.68359375	-11.2780192810167\\
5.7421875	-12.6256894450116\\
5.80078125	-14.545643261883\\
5.859375	-17.3569276335955\\
5.91796875	-21.9257269840793\\
5.9765625	-32.7927948659125\\
6.03515625	-29.3540477221692\\
6.09375	-21.0152721659968\\
6.15234375	-17.0778346054717\\
6.2109375	-14.6328813384305\\
6.26953125	-12.9920404495631\\
6.328125	-11.8849871639844\\
6.38671875	-11.1819139548433\\
6.4453125	-10.8156873612075\\
6.50390625	-10.7536885140371\\
6.5625	-10.9869052961182\\
6.62109375	-11.5272260747623\\
6.6796875	-12.4110287252721\\
6.73828125	-13.711623316777\\
6.796875	-15.5712330809241\\
6.85546875	-18.2914841874765\\
6.9140625	-22.6729501184647\\
6.97265625	-32.5842545435404\\
7.03125	-31.4868548834307\\
7.08984375	-22.474219484623\\
7.1484375	-18.3723192486008\\
7.20703125	-15.8436434936262\\
7.265625	-14.1466174601356\\
7.32421875	-12.995828972503\\
7.3828125	-12.2554359393542\\
7.44140625	-11.855285868919\\
7.5	-11.7609125905568\\
7.55859375	-11.9618804470886\\
7.6171875	-12.4686411512658\\
7.67578125	-13.315676941396\\
7.734375	-14.5731564011286\\
7.79296875	-16.3769377500903\\
7.8515625	-19.0124493407621\\
7.91015625	-23.2212821704827\\
7.96875	-32.3409632321172\\
8.02734375	-33.545538021228\\
8.0859375	-23.7415546870197\\
8.14453125	-19.4675724160151\\
8.203125	-16.8549842704916\\
8.26171875	-15.1032336383824\\
8.3203125	-13.9107113715428\\
8.37890625	-13.1352114255053\\
8.4375	-12.7034410981078\\
8.49609375	-12.5790400753767\\
8.5546875	-12.7501377606443\\
8.61328125	-13.2257642664468\\
8.671875	-14.0385566976354\\
8.73046875	-15.2556670077321\\
8.7890625	-17.0069214897666\\
8.84765625	-19.5626639757578\\
8.90625	-23.6112857470718\\
8.96484375	-32.0616601994606\\
9.0234375	-35.6124409652705\\
9.08203125	-24.8578618386501\\
9.140625	-20.4020271715289\\
9.19921875	-17.7042046157197\\
9.2578125	-15.8982313836596\\
9.31640625	-14.6650826652861\\
9.375	-13.8558332627152\\
9.43359375	-13.3939202332511\\
9.4921875	-13.2410381075285\\
9.55078125	-13.3838671905229\\
9.609375	-13.8300283649247\\
9.66796875	-14.6103602071328\\
9.7265625	-15.7891147528688\\
9.78515625	-17.4903959385262\\
9.84375	-19.9704790389227\\
9.90234375	-23.8696231635535\\
9.9609375	-31.7448286430829\\
10.01953125	-37.7799105679031\\
10.078125	-25.8510490707757\\
10.13671875	-21.2016113517796\\
10.1953125	-18.4164328779364\\
10.25390625	-16.5561164254788\\
10.3125	-15.2828811967257\\
10.37109375	-14.4407001167815\\
10.4296875	-13.9496010658504\\
10.48828125	-13.76927854574\\
10.546875	-13.8849472561466\\
10.60546875	-14.302828998549\\
10.6640625	-15.0520070805887\\
10.72265625	-16.1939430513181\\
10.78125	-17.847306745564\\
10.83984375	-20.2552338860075\\
10.8984375	-24.014310436841\\
10.95703125	-31.3886529039635\\
11.015625	-40.1762200933774\\
11.07421875	-26.7411344938061\\
11.1328125	-21.884324474785\\
11.19140625	-19.0090396782872\\
11.25	-17.0938194222554\\
11.30859375	-15.7806502005327\\
11.3671875	-14.9059905615451\\
11.42578125	-14.3863109795468\\
11.484375	-14.1792472224676\\
11.54296875	-14.2685296841601\\
11.6015625	-14.6589892310481\\
11.66015625	-15.3779941332601\\
11.71875	-16.4843182140936\\
11.77734375	-18.0914659727826\\
11.8359375	-20.4302845991007\\
11.89453125	-24.0576254375828\\
11.953125	-30.9909632800209\\
12.01171875	-43.02013417578\\
12.0703125	-27.542962728271\\
12.12890625	-22.4628148794023\\
12.1875	-19.4941303919279\\
12.24609375	-17.5231093507597\\
12.3046875	-16.1698752744486\\
12.36328125	-15.2629274093804\\
12.421875	-14.7150207473495\\
12.48046875	-14.4816696911094\\
12.5390625	-14.5450994786536\\
12.59765625	-14.9087562998969\\
12.65625	-15.598330802071\\
12.71484375	-16.6700056599856\\
12.7734375	-18.2323708998658\\
12.83203125	-20.5047660577822\\
12.890625	-24.0077989911987\\
12.94921875	-30.5491643686636\\
13.0078125	-46.7828818527665\\
13.06640625	-28.267840902727\\
13.125	-22.9458981224226\\
13.18359375	-19.8800147393656\\
13.2421875	-17.8520187992265\\
13.30078125	-16.458367151479\\
13.359375	-15.5191193735689\\
13.41796875	-14.9431463832581\\
13.4765625	-14.6837745001241\\
13.53515625	-14.7217007972011\\
13.59375	-15.0589912231441\\
13.65234375	-15.7196935587087\\
13.7109375	-16.7574901372876\\
13.76953125	-18.2762908790291\\
13.828125	-20.4846453092903\\
13.88671875	-23.8700226368231\\
13.9453125	-30.0601411272461\\
14.00390625	-52.9458921008178\\
14.0625	-28.9245720793235\\
14.12109375	-23.3394695567303\\
14.1796875	-20.1720881092491\\
14.23828125	-18.085698385969\\
14.296875	-16.6510903690898\\
14.35546875	-15.6793647629028\\
14.4140625	-15.0753285451837\\
14.47265625	-14.7900489451278\\
14.53125	-14.8026696627541\\
14.58984375	-15.1138790539918\\
14.6484375	-15.7461142586641\\
14.70703125	-16.7506426692326\\
14.765625	-18.2269132347748\\
14.82421875	-20.3733459769103\\
14.8828125	-23.6470413479584\\
14.94140625	-29.5201342109041\\
15	-inf\\
};

\end{axis}
\end{tikzpicture}%

%% file: Conclusion.tex
In this work, we presented an OTFS modulation based on the \emph{discrete} Zak transform. We further derived the input-output relation for the symbols in the delay-Doppler domain based on \emph{discrete} Zak transform properties.

OTFS can solely be studied and analyzed by using the Zak transform. The Zak transform perspective allows for a paradigm change: OTFS does not have to be considered an overlay for OFDM. For example, in the OTFS overlay for OFDM, system parameters such as the number of subcarriers are determined by the underlying OFDM system. However, these parameters may not be the optimal choice for OTFS. 

On the other hand, designing an OTFS system based on the Zak transform allows exploring different parameters.  We have shown, for instance, that the frame size directly influences the Doppler resolution. A smaller frame leads to a smaller Doppler resolution which reduces the spread in the Doppler domain. However, a small frame also reduces the spectral efficiency. Thus, finding an optimal trade-off is an interesting research direction.

%% file: Appendices/appendix_DFT_Zak.tex
Substituting $x[n]$ in \eqref{eq:DZT} by \eqref{eq:IDFT}, we obtain
\begin{equation}
    \Zx= \frac{1}{K\sqrt{L}}\sum_{l=0}^{K-1} \sum_{k'=0}^{KL-1} X[k'] e^{j2\pi
        \left(\frac{k'}{KL}(n+lL)-\frac{k}{KL}l\right)}.
\end{equation}
Note that in the derivation of \eqref{eq:IDFT}, the case for $L=1$ was considered, and thus, the sequence $x$ has period $K$. Here, on the other hand, we consider the sequence $x$ to be $KL$-periodic. Therefore, \eqref{eq:IDFT} is adopted accordingly by substituting $K$ by $KL$. Next, we rearrange terms and obtain
\begin{equation}
    \Zx= \frac{1}{K\sqrt{L}} \sum_{k'=0}^{KL-1} X[k']  e^{j2\pi
        \frac{k'}{KL}n}\sum_{l=0}^{K-1} e^{-j2\pi\frac{k'-k}{K}l},
\end{equation}
where we finally replace the last sum by relation \eqref{eq:delta_train} which leads due to the sifting property of the Kronecker delta to 
\begin{equation}
    \Zx=\frac{1}{\sqrt{L}}\sum_{l=0}^{L-1}X[k+lK]e^{j2\pi\frac{k+lK}{KL}n}.
\end{equation}

%% file: Appendices/appendix_inv_DZT.tex
In a first step, we rewrite the summation in \eqref{eq:DFT_1} as a double summation, i.e.,
\begin{equation}
    X[k] = \frac{1}{\sqrt{KL}} \sum_{l=0}^{K-1}\sum_{n=0}^{L-1}x[n+lL]e^{-j\frac{k}{KL}(n+lL)}.
\end{equation}
Next, we use relation \eqref{eq:IDZT2} to express $x[n+lL]$ through its IDZT which leads to
\begin{equation}
 X[k] = \frac{1}{K\sqrt{L}}\sum_{l=0}^{K-1}\sum_{n=0}^{L-1}\sum_{k'=0}^{K-1}Z[n,k'] e^{-j\frac{k-k'}{K}l}e^{-j\frac{k}{KL}n},
\end{equation}
and in a final step we use relation \eqref{eq:delta_train} with respect to the summation over $l$, which results in
\begin{equation}
 X[k]  = \frac{1}{\sqrt{L}}\sum_{n=0}^{L-1} Z_x[n,k]e^{-j\frac{k}{KL}n}.
\end{equation}

%% file: Appendices/appendix_prod.tex
To prove the relation \eqref{eq:DZTproductInv} we substitute the DZT $Z_x$ and $Z_y^{\ast}$ by their definition in \eqref{eq:DZT}. After rearranging terms, we obtain:
\begin{multline}
   \frac{1}{K} \sum_{n=0}^{L-1} \sum_{l'=0}^{L-1}\sum_{l''=0}^{L-1}x[n+l'L]y^{\ast}[n+l''L] e^{-j2\pi\frac{l}{L}n} \\ \sum_{k=0}^{K-1}e^{-j2\pi\frac{k}{K}(l'-l''-m)}.
\end{multline}
We can us relation \eqref{eq:delta_train} to substitute the last summation. By the sifting property of the Kronecker delta \eqref{eq:delta_train}, we further get
\begin{equation}
 \sum_{n=0}^{L-1} \sum_{l'=0}^{L-1}x[n+l'L]y^{\ast}[n+(l'-m)L] e^{-j2\pi\frac{l}{L}n}.
\end{equation}
Since the complex exponential sequence is periodic with period $L$, we can rewrite the double summation as a single summation providing us with 
\begin{equation}
 \sum_{n=0}^{KL-1}x[n]y^{\ast}[n-mL] e^{-j2\pi\frac{l}{L}n} 
\end{equation}
which can be recognized as the inner product between $x$ and $y_{m,l}$.

%% file: Appendices/appendix_mod.tex
To prove the modulation property, use the definition of the sequence $z = x\cdot y$ and the definition of the DZT in \eqref{eq:DZT} which is
\begin{equation}
Z_z[n,k] =\frac{1}{\sqrt{K}} \sum_{l=0}^{K-1}x[n+lL]y[n+lL]e^{-j2\pi\frac{k}{K}l}.
\end{equation}
Now, expressing $x[n+lL]$ using \eqref{eq:IDZT2} we have
\begin{equation}
Z_z[n,k]= \frac{1}{K} \sum_{m=0}^{K-1} Z_x[n,m]  \sum_{l=0}^{K-1}y[n+lL]e^{-j2\pi\frac{(k-m)}{K}l}.
\end{equation}
Finally, we use the DZT definition \eqref{eq:DZT} and obtain
\begin{equation}
    Z_z[n,k]=\frac{1}{\sqrt{K}} \sum_{m=0}^{K-1} Z_x[n,m] Z_y[n,k-m].
\end{equation}

%% file: Appendices/appendix_conv.tex
To prove relation \eqref{eq:convProp}, we first express the circular convolution as a muliplitcation in the DFT domain, i.e., 
\begin{equation}
    Z[k]= \sqrt{KL} X[k]Y[k],
\end{equation}
where the factor $\sqrt{KL}$ is due to the unitary definition of the DFT. Using \eqref{eq:Janssen}, we get
\begin{equation}
    Z_z[n,k] = \sqrt{K} \sum_{l=0}^{L-1} X[k+lK]Y[k+lK]e^{j2\pi\frac{k+lK}{KL}n}.
\end{equation}
Now, using \eqref{eq:IDZTfreq} to express the elements of the DFT through their DZT we obtain
\begin{multline}
     Z_z[n,k] = \frac{\sqrt{K}}{L} \sum_{n'=0}^{L-1} \sum_{n''=0}^{L-1}Z_x[n',k] Z_y[n'',k] \\ \sum_{l=0}^{L-1} e^{-j2\pi\frac{k+lK}{KL}(n'+n''-n)}.
\end{multline}
Substituting the last sum by \eqref{eq:delta_train} and applying the sifting property of the Kronecker delta, we finally get
\begin{equation}
     Z_z[n,k] =   \sqrt{K} \sum_{n'=0}^{L-1}Z_x[n',k] Z_y[n-n',k].
\end{equation}


%% file: acknowledgment.tex
The authors would like to thank Dr. Recep Firat Tigrek (ASML) and Dr. Alessio Filippi (NXP Semiconductors) for fruitful discussions on OTFS.